\documentstyle[12pt,supercite]{article}
\newcommand{\Obs}{{\cal O}}
\newcommand{\eps}{\epsilon}

\newcommand{\Ubar}{\overline{U}}
\newcommand{\la}{\langle}
\newcommand{\ra}{\rangle}
\newcommand{\tint}{\tau_{\rm int}}
\newcommand{\texp}{\tau_{\rm exp}}

\newlength{\digitwidth} 

\def\sqr#1#2{{\vcenter{\hrule height.#2pt
      \hbox{\vrule width.#2pt height#1pt \kern#1pt
        \vrule width.#2pt}
      \hrule height.#2pt}}}

\def\abstracts#1#2#3{{
        \centering{\begin{minipage}{4.62in}\baselineskip=13pt
        \small
        \centerline{\bf Abstract}
        \vspace*{0.2cm}                
        \parindent=0pt #1\par
        \parindent=18pt #2\par
        \parindent=15pt #3
        \end{minipage} }\par}}

\renewcommand{\thefootnote}{\fnsymbol{footnote}}
\begin{document}
\vspace*{-3cm}
\hfill \parbox{4.2cm}{ 
                       FUB-HEP 01/97\\
                revised June 1997 \\
               }\\[0.8cm]
%
\vspace*{0.1cm}
\centerline{\LARGE \bf Optimal Energy Estimation } \\[0.2cm]
\centerline{\LARGE \bf in Path-Integral Monte Carlo } \\[0.2cm]
\centerline{\LARGE \bf Simulations}\\[0.8cm]
%
\renewcommand{\thefootnote}{\arabic{footnote}}
\vspace*{0.3cm}
\centerline{\large {\em Wolfhard Janke\/}$^{1,2}$ 
               and {\em Tilman Sauer\/}$^{2,3,\star}$}\\[0.4cm] 
\centerline{\large $^1$ {\small Institut f\"ur Physik,
                        Johannes Gutenberg-Universit\"at Mainz}}
\centerline{            {\small 55099 Mainz, Germany }}
\centerline{email: \verbß janke@miro.physik.uni-mainz.de ß}\\[0.15cm]
\centerline{\large $^2$ {\small Institut f\"{u}r Theoretische Physik,
                        Freie Universit\"{a}t Berlin}}
\centerline{            {\small 14195 Berlin, Germany}}\\[0.15cm]
\centerline{\large $^3$ {\small Max-Planck-Institut f\"ur
                        Wissenschaftsgeschichte}}
\centerline{            {\small Wilhelmstra{\ss}e 44, 10117 Berlin, 
                        Germany}}\\[0.15cm]
\centerline{\large $\star$ {\small present address:}}
\centerline{ {\small  Institut f\"ur Wissenschaftsgeschichte,
                        Georg-August-Universit\"at G\"ottingen}}
\centerline{            {\small Humboldtallee 11, 37073 G\"ottingen, 
                        Germany}}
\centerline{email: \verbß tsauer@gwdg.de ß}\\[2.50cm]
%

%
                    \abstracts{}{ We investigate the properties of two
                      standard energy estimators used in path-integral
                      Monte Carlo simulations. By disentangling the
                      variance of the estimators and their
                      autocorrelation times we analyse the dependence
                      of the performance on the update algorithm and
                      present a detailed comparison of refined update
                      schemes such as multigrid and staging
                      techniques.  We show that a proper combination
                      of the two estimators leads to a further
                      reduction of the statistical error of the
                      estimated energy with respect to the better of
                      the two without extra cost.  }{}
  \thispagestyle{empty}
    \newpage
      \pagenumbering{arabic}
%
             \section{Introduction}                           \label{sec:intro}
%
A detailed understanding of the statistical properties of
many-particle quantum systems is among the most challenging objectives
in condensed matter physics and physical chemistry. Apart from a few
simple model systems, analytical approaches can usually only provide
an approximative description whose accuracy is difficult to control
inherently. Computer simulations, on the other hand, in principle yield
exact results even for complicated systems such as, for
example, non-linear quantum chains or quantum crystals. A combination
of the two complementary approaches may thus result in valuable
insights into the physics of quantum systems.

While in principle an exact method, the difficulty of computer
simulations is to actually achieve the desired accuracy in
practice. For path-integral Monte Carlo (PIMC) methods, which we shall
consider in this paper, the draw-backs are well known \cite{mc}. 
Being a stochastic method, all results are subject to statistical errors
which, in principle, can be made as small as desired by increasing the 
simulation time. 
The necessary
discretization of the path integral, however, requires an extrapolation to the
continuum limit where standard local update algorithms exhibit a
severe slowing down. By this one means that successively generated
paths, or more generally, configurations of a many-particle system,
are highly correlated in the Monte Carlo process. This effect greatly
enhances the statistical errors of the measurements in a given
computer time.

More precisely the statistical uncertainty for 
the mean value $\overline{\cal O} = (1/N_m)$ $\sum_{i=1}^{N_m} {\cal O}_i$ of
an observable 
${\cal O} = \la \overline{\cal O} \ra$ 
measured in an importance sampling Monte Carlo process is given, in general,
by the error estimate 
\begin{equation}
 \label{eq:errorformula}
      \Delta \overline{\cal O} = \sqrt{\frac{\sigma_{{\cal O}_i}^2}{N_m}}
               \sqrt{2\tint}\,,
\end{equation}
where $\sigma_{{\cal O}_i}^2$ is the variance of the
individual autocorrelated measurements ${\cal O}_i$ at ``time'' $i$, 
$\tint$ is the integrated autocorrelation time,      
and $N_m$ is the total number of measurements.
As it appears, there are three ways to reduce the statistical 
error. The most obvious but most expensive one is to increase the number
of measurements. Since, however, $1/\sqrt{N_m}$ is a rather slowly
decreasing function and, in the physically interesting continuum
limit, it may happen that both the variance and the autocorrelation time
diverge with some power of the discretization parameter, it is much more
promising to ask whether the statistical error may also be reduced by 
constructing an estimator for ${\cal O}$ with a smaller variance or by 
employing refined update algorithms with smaller autocorrelation times.
As we shall see below, the latter two strategies are to some extent 
intertwined, a fact which calls for a careful analysis.

In this paper we will focus on energy estimation.  It is well
known that in path-integral Monte Carlo simulations the energy may be
measured using two different estimators.  One is derived by a
straightforward differentiation of the partition function and will be
called for the sake of brevity the ``kinetic'' estimator since it
involves an explicit measurement of the kinetic part of the energy
\cite{barker79}. The other is based on the virial theorem for path
integrals and will hence be referred to as the ``virial'' estimator
\cite{cf81,hbb82}. The ``virial'' estimator is often judged to be the
``better'' estimator because in the continuum limit its variance is
much smaller than that of the ``kinetic'' estimator.

Early investigations of the ``kinetic'' and the ``virial'' estimators
focussed on their variances~\cite{hbb82,pr84}.  In the following
years it was pointed out that a correct assessment of the accuracy
also has to take into account the autocorrelations, and it was
demonstrated that for a standard Metropolis simulation of the harmonic
oscillator the allegedly less successful ``kinetic'' estimator gave
smaller errors than the ``virial'' estimator \cite{gj88}. In a further
investigation~\cite{cb89} it was shown that conclusions about the
accuracy also depend on the particular Monte Carlo update algorithm at
hand since modifications of the update scheme such as inclusion of
collective moves of the whole path affect the autocorrelations of the
two estimators in a different way.

For a fair judgement of the performance of an estimator, one thus has
to take into account both the variances and the associated
autocorrelation times. In this paper we analyze this problem for the
two standard estimators in combination with four different update
algorithms, namely (1) the standard Metropolis algorithm, 
(2) the multigrid V-cycle, (3) the multigrid W-cycle, and (4) the staging
algorithm. The Metropolis algorithm is based on {\em local\/} moves and 
exhibits severe slowing down in the continuum limit
\cite{js92b,js93a}. The other three update algorithms employ {\em non-local\/} 
moves and therefore reduce (V-cycle) or even completely overcome (W-cycle, 
staging) the slowing-down problem \cite{js93a,js96a}. 
As our main result we will then demonstrate how the estimation of energy can be
improved at practically no extra cost by taking a suitable linear combination
of the two estimators. We shall see that the optimal combination has to take
into account not only the variances of the two individual estimators but also
their covariance, i.e., the cross-correlations between the two estimators.

The outline of the paper is as follows. In the next section we first recall the
definition of the kinetic and virial estimators for the energy and
discuss some of their basic properties. We then introduce the ``combined''
estimator and present a theoretical analysis of its properties. 
In Sec.~3 we define autocorrelation times and describe
how our error analysis was performed.
The various
update algorithms are briefly summarized in Sec.~4. 
The results of the numerical
simulations are contained in Sec.~5, and in Sec.~6 we close with our
conclusions and a few final remarks. 
%
          \section{Energy estimators}                    \label{sec:estimators}
%
For didactic reasons we shall illustrate the improved energy estimation
for simple
one-particle quantum systems governed by a Hamiltonian~\cite{fn1}
\begin{equation}
\hat{H} = \frac{1}{2m}\hat{p}^2 + V(\hat{x}).
\label{eq:hamilton}
\end{equation}
where $m$ is the mass of the particle and $V(x)$ a potential to be
specified below.  Our theoretical considerations are, however,
completely general and without any additional problems applicable to
many-body quantum systems as well.
When coupled to a heat-bath at inverse
temperature $\beta = 1/k_B T$, the canonical partition function is given
in the operator representation by
\begin{equation}
{\cal Z}(\beta) = {\rm Tr} e^{-\beta \hat{H}} = \sum_n e^{-\beta E_n},
\label{eq:1}
\end{equation}
where $E_n$ are the energy eigenvalues associated with
(\ref{eq:hamilton}). The equivalent path-integral
representation~\cite{kleinert_1990} we wish to simulate reads
($\hbar\equiv 1$)
\begin{equation}
{\cal Z}(\beta) = \int {\cal D}x \exp\left\{-\int_0^{\beta} d\tau\left[
\frac{m}{2} \dot{x}^2 + V(x(\tau))\right]\right\} = \lim_{L\rightarrow \infty}
{\cal Z}^{(L)}(\beta),
\label{eq:Z_PI}
\end{equation}
where ${\cal Z}^{(L)}$ is the discretized path integral,
\begin{equation}
\!{\cal Z}^{(L)}(\beta) = \left[ \prod_{k=1}^{L} \int_{-\infty}^{\infty}\!\!
\frac{dx_k}{\sqrt{2\pi\eps/m}} \right]
\exp\left\{-\sum_{k=1}^L \left[ \frac{m}{2\eps} \left( x_k-x_{k-1}
\right)^2 + \eps V(x_k)  \right] \right\},
\label{eq:Z_PI_dis}
\end{equation}
with $\epsilon = \beta/L$ being the usual discretization parameter. 
The trace in (\ref{eq:1}) implies periodic boundary conditions,
i.e., in (\ref{eq:Z_PI_dis}) we take $x_0 = x_L$.

For the potential $V(x)$ we chose two characteristic shapes covering a wide
range of physical phenomena. The first one is an anharmonic convex potential,
\begin{equation}
        V(x) = 0.5x^2 + x^4 \qquad {\rm (CP)},
\label{eq:CP}
\end{equation}
relevant for studying fluctuations around a unique minimum,
and the second one is a double-well potential,
\begin{equation}
        V(x) = -0.5x^2 + 0.04 x^4 \qquad {\rm (DW)},
\label{eq:DW}
\end{equation}
which exhibits tunneling phenomena. 

\subsection{The ``kinetic'' estimator}         \label{ssec:Kinetic}

From the partition function (\ref{eq:1}) it is clear that the average
energy is defined by
\begin{equation}   
U = \la \hat{H} \ra = -\frac{\partial \ln {\cal Z}}{\partial \beta} = 
{\rm Tr} \hat{H} e^{-\beta \hat{H}}/{\cal Z} =
\frac{\sum_n E_n e^{-\beta E_n}}{\sum_n e^{-\beta E_n}}.
\end{equation}
Since $\partial/\partial \beta = (1/L) \partial / \partial \epsilon$
it is easy to see that we get, in the 
path-integral representation,
$U = \lim_{L\rightarrow \infty} U^{(L)}$ with $U^{(L)} = \la U_{\rm k} \ra$
and 
\begin{equation}   
U_{\rm k} = \frac{L}{2\beta} - \frac{m}{2L} \sum_{k=1}^L 
\left( \frac{x_k - x_{k-1}}{\epsilon} \right)^2 + 
\frac{1}{L} \sum_{k=1}^L V(x_k), 
\label{eq:kineticestimator}
\end{equation}
and $\la \dots \ra$ now means expectation values with respect to the
discretized path integral (\ref{eq:Z_PI_dis}), which are therefore
also $L$-dependent. The term $L/2\beta$ stems from the
$\beta$-dependence of the measure.
For the sake of brevity we will in the sequel refer to the energy
estimator (\ref{eq:kineticestimator}) as the {\em kinetic}
estimator~\cite{barker79} of the energy identified by a subscript
${\rm k}$ since it involves an explicit measurement of $\la p^2/2m \ra
= \lim_{L\rightarrow\infty}\la T_{\rm k}\ra^{(L)}$, with
\begin{equation}
T_{\rm k} \equiv \frac{L}{2\beta} - \frac{m}{2L}
 \sum_{k=1}^{L} \left( \frac{x_k - x_{k-1}}{\eps} \right)^2 
\label{eq:10}
\end{equation}
being the path-integral estimator of the kinetic part of the energy.
Notice that the kinetic part is {\em not\/} given by 
\begin{equation}
\frac{m}{2} \la v_k^2 \ra
\equiv \frac{m}{2} \la\frac{1}{L} \sum_{k=1}^L 
\left( \frac{x_k - x_{k-1}}{\eps} \right)^2\ra,
\end{equation}
as one might have naively guessed,
since for all reasonably well
behaved potentials 
$m \la v_k^2 \ra/2 = L/2\beta + {\cal O}(1)$
diverges as $L \longrightarrow \infty$ (and $\beta$ fixed as usual). For the
proper definition, however, this piece is just needed to cancel the divergence
coming from the measure.

Numerically it appears questionable,
however, at first sight whether this estimator is very useful. The
reason is that most of the signal of $\la (x_k - x_{k-1})^2 \ra$ is
used to cancel the explicit $L/2\beta$ term.  We would therefore
expect that the variance of the ``kinetic'' estimator,
\begin{equation}
\sigma^2_{\rm k} = \la U^2_{\rm k} \ra - \la U_{\rm k} \ra^2,
\label{eq:sigma_def}
\end{equation}
diverges linearly with $L$, and we would hence expect large statistical errors
in the continuum limit of the path integral even for update algorithms
which overcome the slowing down problem mentioned in the Introduction.
This is what we indeed observed empirically; see the discussion in
section~\ref{sec:results} below.
What we need is another estimator $\tilde{T}$ for the kinetic energy 
with the {\em same mean}, 
$\la \tilde{T} \ra = \la \hat{p}^2/2m \ra$, but {\em smaller variance\/}. 
In the statistical literature this
would run under the label of a ``variance reduced'' estimator. One
possible candidate is suggested by means of the path-integral version
of the {\em virial theorem\/}.

\subsection{The ``virial'' estimator}             \label{ssec:Virial}

In operator language the virial theorem states that~\cite{merzbacher}
\begin{equation}
\la  \frac{ \hat{p}^2} {2m} \ra = \frac{1}{2} \la \hat{x} V'(\hat{x}) \ra
\label{eq:12}
\end{equation}
where we have used the abbreviation $V'(x) = dV/dx$. From the above discussion
one may guess that the path-integral analog should read
\begin{equation}
\frac{L}{2\beta} - \frac{m}{2} \la \left( \frac{x_k - x_{k-1}} {\eps} \right)^2
\ra = \frac{1}{2} \la x_k V'(x_k) \ra .
\label{eq:13}
\end{equation}
This relation indeed follows from the
invariance of the partition function under a rescaling
of the ``field'' $x_k$. 
To show this, we start from eq.~(\ref{eq:Z_PI_dis}) and
rescale the ``field'' $x_k \longrightarrow \lambda x_k$, which gives
\begin{equation}
{\cal Z}^{(L)} = \left[ 
\prod_{i=1}^{L} \int_{-\infty}^{\infty}
\frac{dx_k}{\sqrt{\frac{2\pi\eps}{m\lambda^2}}} \right]
\exp\left\{ -\sum_{i=1}^L \left[ \frac{m}{2\eps} \lambda^2 \left( x_k-x_{k-1}
\right)^2 + \eps V(\lambda x_k) \right] \right\} .
\label{eq:15}
\end{equation}
By choosing $\lambda^2\equiv\eps/m$ the factor in front of the
kinetic term becomes $1/2$ and also the $\beta$-dependence of the measure
vanishes. Differentiation with respect to $\beta$ now only 
affects the potential energy term.
Noting that
$\partial V(\lambda x)/\partial \lambda = x \partial V(y)/
\partial y |_{y=\lambda x}$,
and rescaling variables back after differentiation, one readily arrives at
$U^{(L)} = \la U_{\rm v} \ra$ with
the alternative ``virial'' estimator~\cite{cf81,hbb82}
\begin{equation}
U_{\rm v} = \frac{1}{2L} \sum_{k=1}^L 
x_k V'(x_k) + \frac{1}{L} \sum_{k=1}^L V(x_k). 
\label{eq:virialestimator}
\end{equation}
From here we deduce the virial theorem (\ref{eq:13}) 
with the well-behaved kinetic energy estimator 
$T_{\rm v} = \frac{1}{2L} \sum_{k=1}^L x_k V'(x_k)$.
The subscript ${\rm v}$ is a reminder of the fact that this energy
estimator was derived using the virial theorem or, alternatively,
that, in contrast to the ``kinetic'' estimator, it only refers to the
potential $V(x_k)$.  An alternative derivation of this estimator which
makes use of the very same rescaling invariance would again start by
rescaling the ``field'' $x_k$ in the partition function
(\ref{eq:Z_PI_dis}) by $x_k \rightarrow \lambda x_k$. Differentiating
$\ln {\cal Z}^{(L)}$ with respect to $\lambda$ and putting $\lambda =
1$, one obtains in fact directly relation (\ref{eq:13}). The virial
estimator can also be derived in yet another way \cite{hbb82}.

Notice that using the (incorrect) estimator $\frac{m}{2} \la v_k^2
\ra$, identifying it (incorrectly) with $\la \frac{ \hat{p}^2 }{2m}
\ra$, applying the (correct) identity (operator formulation of the
virial theorem), $\la \frac{ \hat{p}^2}{2m} \ra = \la \frac{1}{2}
\hat{x} V'(\hat{x}) \ra$, and identifying $\la \frac{1}{2} \hat{x}
V'(\hat{x}) \ra$ (correctly) with $\la \frac{1}{2} x_k V'(x_k) \ra$
one arrives (accidentally) at the same result, but the derivation is
clearly dubious.

\subsection{Discussion of the two estimators}  \label{ssec:Discuss}

As explained above, the variance of the measured energies is a 
property of the estimators we employ. Regarding the dependence on the
discretization, we expect asymptotically for large $L$
a linear divergence of the variance for the
``kinetic'' estimator whereas the variance should stay roughly
constant for the ``virial'' estimator. We emphasize that this dependence
is completely independent of the update algorithm.

\begin{figure}[tb]
\vskip 9.0truecm
\includegraphics{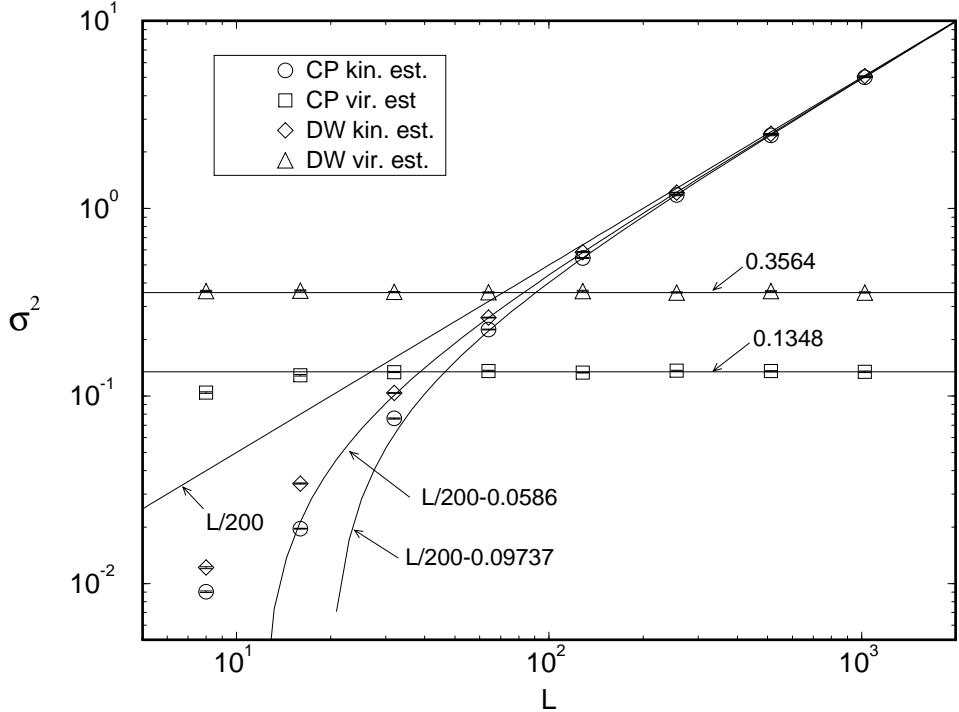}
\caption[a]{
  Variance of the individual energy measurements using the ``kinetic''
  and the ``virial'' estimators for the two potentials (\ref{eq:CP}) and
  (\ref{eq:DW}) at $\beta=10$. While the variance
  of the virial estimator is roughly constant in the continuum limit
  $L \longrightarrow \infty$, the variance of the
  kinetic estimator asymptotically diverges as $\sigma_{\rm k}^2 = 
  L/2\beta^2$.}
\label{figure:sigma}
\end{figure}
For purposes of illustration we show in Fig.~\ref{figure:sigma} the
variance of the two estimators as a function of the number of
variables $L$. Since the update algorithm only affects the
autocorrelation times and is {\em a priori\/} irrelevant for the
variance of the individual measurements we may choose data for any
update algorithm. In Fig.~\ref{figure:sigma} we have used
the data obtained by the multigrid W-cycle, cf.\ the discussion in
sections \ref{ssec:Multi} and \ref{ssec:Simu} and Table
\ref{table:cptau} below. As expected, the measured variances for the
virial estimator are indeed roughly constant for all values of $L$.
Minor deviations from the asymptotic value of $\sigma_{\rm v}^2 =
0.1348(12)$ (CP) resp. $0.3564(24)$ (DW) (cp.  Table
\ref{table:cptau}) are observed only for the coarsest discretization
of $L=8$.
 
The behaviour is clearly different for the kinetic energy estimator.
We first observe that for both potentials the divergence turns indeed
out to be linear for large enough values of $L$. The straight line in
Fig.~\ref{figure:sigma} is the expected asymptotic behaviour of
$\sigma_{\rm k}^2 \rightarrow L/2\beta^2$, which in fact fits the data
well for $L = 256$, $512$, and $1024$. In order to account for the
discrepancies which are observed for smaller values of $L$, we looked
at the first correction term for $\sigma_{\rm k}^2$. A straightforward
calculation shows that the variance (\ref{eq:sigma_def}) of the
kinetic estimator is given quite generally by
\begin{equation}
 \sigma_{\rm k}^2 = \frac{L}{2\beta^2} + \frac{C}{\beta^2}
                    - \frac{1}{\beta}\langle x_iV'(x_i)\rangle,
\end{equation}
where $C = -\beta^2 \partial U/\partial \beta$ is the specific heat. 
Instead of evaluating $C$ directly
we first calculate the variance of the virial estimator,
\begin{equation}
\sigma_{\rm v}^2 = \frac{C}{\beta^2} + \frac{1}{\beta}
\langle \frac{3}{4}x_iV'(x_i)+\frac{1}{4}x_i^2V''(x_i)\rangle
\end{equation}
which allows us to express $\sigma_{\rm k}^2$ as
\begin{equation}
\label{eq:sig_k}
\sigma_{\rm k}^2 = \frac{L}{2\beta^2} + \sigma_{\rm v}^2 -
\frac{1}{\beta}
\langle \frac{7}{4}x_iV'(x_i)+\frac{1}{4}x_i^2V''(x_i)\rangle.
\end{equation}
For the quartic potential $V(x)
= \omega^2x^2/2+gx^4$ we then find
\begin{equation}
\langle \frac{7}{4}x_iV'(x_i)+\frac{1}{4}x_i^2V''(x_i)\rangle
    = 2\omega^2\langle x_i^2\rangle + 10g\langle x_i^4 \rangle.
\end{equation}
The expectation value on the right hand side can be expanded as
$a_0 + a_1/L^2 + a_2/L^4 + \dots$.
In order to compute the first correction term $a_0$ we now use  
expectation values for the moments for $L=128$, assuming that the
$1/L^2$ corrections are already negligible for this value of $L$.
At $\beta=10$ we obtained (in a separate simulation using the staging 
algorithm)
$\langle x_i^2 \rangle = 0.25660(46)$
and $\langle x_i^4 \rangle = 0.18085(62)$ for the convex potential
(CP, $\omega^2=1$, $g=1$), resp. $5.391(18)$ and $37.33(19)$ for the 
double-well potential (DW, $\omega^2=-1$, $g=0.04$). Numerically, we therefore
find the correction terms to be $a_0 = 2.3217$ (CP) resp. $4.1500$ (DW), and
together with the asymptotic values of $\sigma^2_{\rm v} = 0.1348$ (CP)
resp. $0.3564$ (DW) we finally arrive at 
$\sigma^2_{\rm k} - L/2\beta^2 = -0.09737$ (CP) resp. $-0.0586$ (DW).  
As can be seen from Fig.~\ref{figure:sigma}, taking into account this
first correction does indeed reproduce the data down to at least $L=64$.
Clearly, for smaller values of $L$ we would need to evaluate the higher-order
coefficients $a_1$, $a_2$, etc.,
in order to reproduce the values for $\sigma_{\rm k}^2$.

In summary, we have illustrated the expected behaviour that the
variances of the measurements are roughly independent from $L$ for the
virial estimator but diverge linearly for large values of $L$ for the
kinetic estimator.  Before going on we finally remark that the two
different energy estimators also entail two different estimators for
the specific heat by virtue of the relation
\begin{equation}
     \la C_{\rm k,v} \ra = - \beta^2 \frac{\partial\la U_{\rm k,v} \ra}
                        {\partial \beta}.
\end{equation}

\subsection{The ``combined'' estimator}             \label{ssec:Combi}

In the Monte Carlo process we are measuring the energy using either
the ``kinetic'' estimator (\ref{eq:kineticestimator}) or the 
``virial'' estimator (\ref{eq:virialestimator}) for the configurations
$\{x_j\}_i$. This procedure yields the two mean values 
$\Ubar_{\rm k} \equiv \sum_{i=1}^{N_m} U_{{\rm k},i}/N_m$ 
and $\Ubar_{\rm v} \equiv \sum_{i=1}^{N_m} U_{{\rm v},i}/N_m$ 
as stochastic variables with the same expectation value,
$\la \Ubar_{\rm k} \ra = \la \Ubar_{\rm v} \ra = U$, but different 
statistical errors. Let us here abbreviate the {\em variances of the mean 
values\/} (i.e., the squared errors) by $\Delta^2_{\rm k} \equiv
(\Delta \Ubar_{\rm k})^2 = 
\la \Ubar_{\rm k}^2 \ra - \la \Ubar_{\rm k} \ra^2$ and $\Delta^2_{\rm v} 
\equiv (\Delta \Ubar_{\rm v})^2 = 
\la \Ubar_{\rm v}^2 \ra - \la \Ubar_{\rm v} \ra^2$. 

It is clear that any linear combination of the form
\begin{equation}
      \Ubar_{\rm c}(\alpha) = \alpha \Ubar_{\rm k} + (1-\alpha) \Ubar_{\rm v}
 \label{eq:linearcombi}
\end{equation}
would also give the same expectation value, $\la \Ubar_{\rm c} \ra = U$, as
the individual estimators alone. The variance
$\Delta^2_{\rm c} \equiv (\Delta \Ubar_{\rm c})^2$
of the combined 
estimator $\Ubar_{\rm c}$ would be given by
\begin{eqnarray}
     \Delta_{\rm c}^2 &=& \la (\alpha \Ubar_{\rm k} 
                         + (1-\alpha) \Ubar_{\rm v})^2\ra
                         - \la  \alpha \Ubar_{\rm k} 
                         + (1-\alpha) \Ubar_{\rm v} \ra^2
                                  \nonumber \\
              &=& \alpha^2 \Delta_{\rm k}^2 + 
                  2\alpha(1-\alpha)\Delta_{\rm kv}^2 +
                  (1-\alpha)^2\Delta_{\rm v}^2,
\label{eq:sigmacombinedgeneral}
\end{eqnarray}
where $\Delta_{\rm kv}^2 = \la \Ubar_{\rm k} \Ubar_{\rm v} \ra
- \la \Ubar_{\rm k} \ra \la \Ubar_{\rm v} \ra$ is the covariance of the 
mean values of the two estimators. Minimizing the variance $\Delta_{\rm c}^2$
with respect to the parameter $\alpha$
yields the optimal $\alpha$ as 
\begin{equation}
     \alpha_{\rm opt} = \frac{\Delta_{\rm v}^2 - \Delta_{\rm kv}^2}
                   {\Delta_{\rm k}^2 + \Delta_{\rm v}^2 - 2\Delta_{\rm kv}^2}.
\label{eq:alphaopti}
\end{equation}
Inserting this optimal $\alpha_{\rm opt}$ into 
eq.~(\ref{eq:sigmacombinedgeneral}) one obtains an expression for the 
minimal variance of $\Ubar_{\rm c}(\alpha)$ which reads
\begin{eqnarray}
     \Delta_{\rm c,opt}^2 &=&
             \frac{\Delta_{\rm k}^2\Delta_{\rm v}^2-(\Delta_{\rm kv}^2)^2}
                  {\Delta_{\rm k}^2 + \Delta_{\rm v}^2 - 2\Delta_{\rm kv}^2}
                        \nonumber \\
              &=& 
             \frac{1-\rho^2}{ 1/\Delta_{\rm k}^2+
                              1/\Delta_{\rm v}^2-
                              2\rho/\Delta_{\rm k}\Delta_{\rm v} } ,
 \label{eq:sigmacombined}
\end{eqnarray}
where $\rho=\Delta^2_{\rm kv}/(\Delta_{\rm k}\Delta_{\rm v})$ is the
correlation coefficient of $\Ubar_{\rm k}$ and $\Ubar_{\rm v}$.
If the two estimators were completely decorrelated, $\rho=0$,
the optimal combination of the two estimators would simply be the error
weighted mean of the individual estimators.
If the variances $\Delta_{\rm k}^2$ and $\Delta_{\rm v}^2$ furthermore 
were equal, the error $\Delta_{\rm c}$ would be reduced
by a factor of $\sqrt{2}$, i.e., one would effectively gain a factor of 
$2$ in the run time. If on the other hand the two variances were very
much different we would not gain very much, since the less accurate
estimator would hardly get any weight in (\ref{eq:linearcombi}) (i.e.,
$\alpha \approx 0$ or $\alpha \approx 1$).

The optimal combination of the two estimators involves, however, the 
covariance of the two estimators and therefore is in general not
simply given as the error weighted mean of the individual estimators.
In fact, erroneously ignoring the covariance of the two estimators might 
lead to the irritating result that the variance of the combined estimator 
is not even reduced with respect to the smaller one of the two individual
estimators.

As it turns out, taking properly into account the covariance of the 
two estimators it may even happen that $\alpha > 1$ or $\alpha<0$, and
it may also happen that the error of the combined estimator is 
even {\em more\/} reduced than by the factor of $\sqrt{2}$, which might 
naively be expected to be an upper bound for the gain in accuracy. 

\begin{figure}[ht]
\unitlength1cm
\begin{minipage}[t]{13.2cm}
\vskip 8.0truecm
\includegraphics{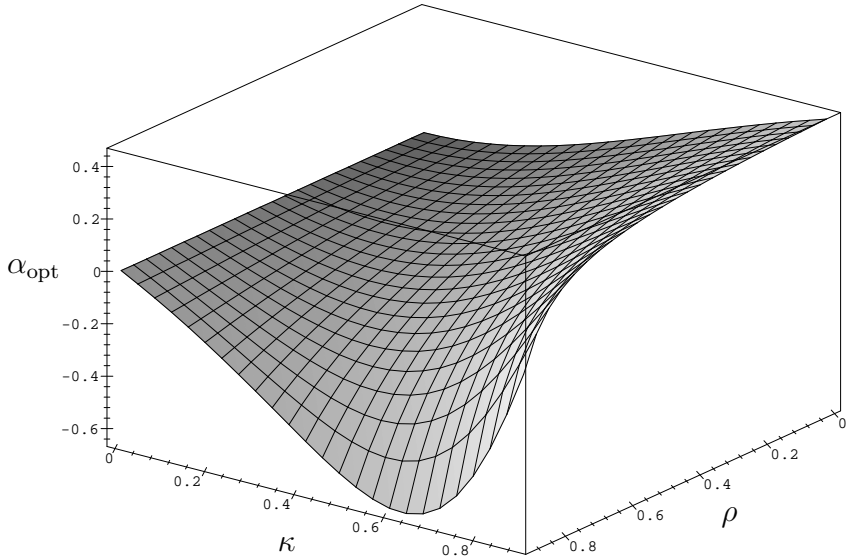}
\begin{picture}(13,2)
\put(0.5,6.5){$\alpha_{\rm opt}$}
\put(4.1,2.8){$\kappa$}
\put(10.0,3.2){$\rho$}
\end{picture}
\end{minipage}
\vspace*{-1.8cm}
\caption[a]{
  The optimal interpolation parameter $\alpha_{\rm opt}$ as given in
  eq.~(\ref{eq:alpha_kappa}) as a function of the ratio 
$\kappa = \Delta_{\rm v}/\Delta_{\rm k}$ and the correlation coefficient 
$\rho=\Delta_{\rm kv}^2/(\Delta_{\rm k} \Delta_{\rm v})$. }
\label{figure:alpha}
\end{figure}

\begin{figure}[ht]
\unitlength1cm
\begin{minipage}[t]{13.2cm}
\vskip 8.0truecm
\includegraphics{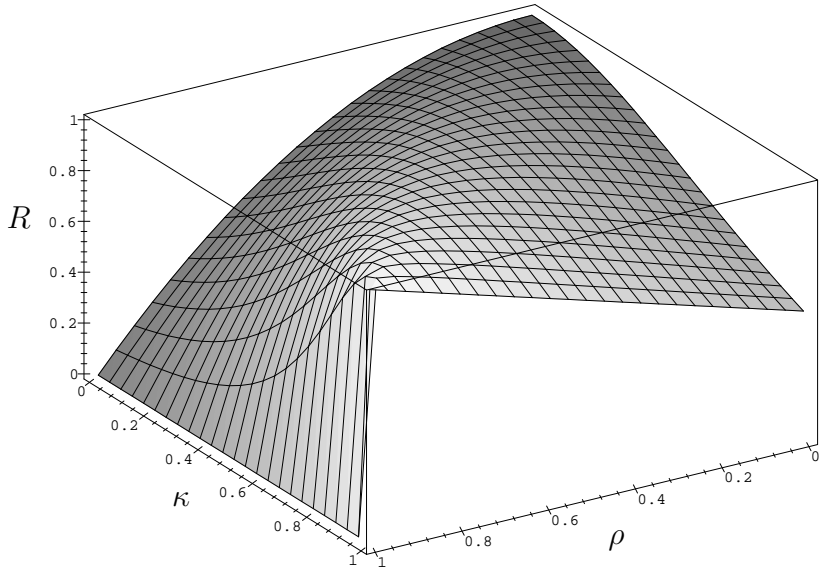}
\begin{picture}(13,2)
\put(0.8,7.1){$R$}
\put(3.0,3.4){$\kappa$}
\put(8.8,2.9){$\rho$}
\end{picture}
\end{minipage}
\vspace*{-1.9cm}
\caption[a]{
    The reduction factor $R$ as given in eq.~(\ref{eq:R}) as a function
of the ratio 
$\kappa = \Delta_{\rm v}/\Delta_{\rm k}$ and the correlation coefficient 
$\rho=\Delta_{\rm kv}^2/(\Delta_{\rm k} \Delta_{\rm v})$. }
\label{figure:R}
\end{figure}

For further discussion, let us assume without loss of generality that 
$\Delta_{\rm v} \le \Delta_{\rm k}$ and let us introduce the ratio
$\kappa = \Delta_{\rm v}/\Delta_{\rm k} \le 1$.
The optimal parameter $\alpha_{\rm opt}$ may then be expressed as
\begin{equation}
 \alpha_{\rm opt} = \frac{\kappa^2-\rho\kappa} {1+\kappa^2-2\rho\kappa},
  \label{eq:alpha_kappa}
\end{equation}
and the reduction of the error may be judged by looking at the quantity
\begin{equation}
  R \equiv \frac{\Delta_{\rm c,opt}^2}{\Delta_{\rm v}^2} 
    =      \frac{1-\rho^2}{1+\kappa^2-2\rho\kappa},
    \label{eq:R}
\end{equation}
which by definition satisfies $0 \le R \le 1$. The smaller $R$, the
greater the gain by using the combined estimator.
For an overview of possible situations, we plot $\alpha_{\rm opt}$
and $R$ as a function of the ratio $\kappa$ and the correlation
coefficient $\rho$ in Figs.~\ref{figure:alpha} and \ref{figure:R}.

In Fig.~\ref{figure:alpha} we first notice that for 
vanishing covariance, $\rho=0$,
the optimal interpolation parameter $\alpha_{\rm opt}$ is always in the range
$0\leq\alpha_{\rm opt}\leq 1$. For completely correlated data, $\rho=1$,
on the other hand, the optimal interpolation parameter
is always negative, $\alpha_{\rm opt} = \kappa/(\kappa - 1) < 0$.

Taking a look at the error reduction factor $R$ in Fig.~\ref{figure:R} 
we notice again, that
for completely 
decorrelated data
$\rho=0$, nothing spectacular
happens. The gain is best if the variances of the two estimators
are equal, $\kappa=1$, and there is no gain at all, if the smaller
variance is negligible compared to the larger one, $\kappa=0$.
The situation is very different though for highly correlated data,
$\rho\rightarrow1$. We first observe that in this case the gain can
be much more profitable than the best gain of $R=0.5$ for completely
decorrelated data. Assuming, for example, a correlation of $\rho=0.9$ 
and a ratio
of $\kappa=0.5$ we find that the reduction factor is $R\approx 0.21$, 
i.e. the variance of the combined estimator is then roughly 5 times
smaller than the smaller of the individual variances. 
It should be noted that for cross-correlated data there is in fact no
limit to the gain in efficiency, and that, as a general rule,
the gain is largest for two estimators which are highly correlated and 
have very different variances. Notice that for $\rho \approx 0$ the inverse
gain factor $R$ is a decreasing function of $\kappa$ (highest gain for
comparable variances), while for $\rho \approx 1$ the situation is just
reversed, i.e., $R$ increases with $\kappa$ (highest gain for very different
variances).
 
\begin{figure}[tb]
\vskip 8.0truecm
\includegraphics{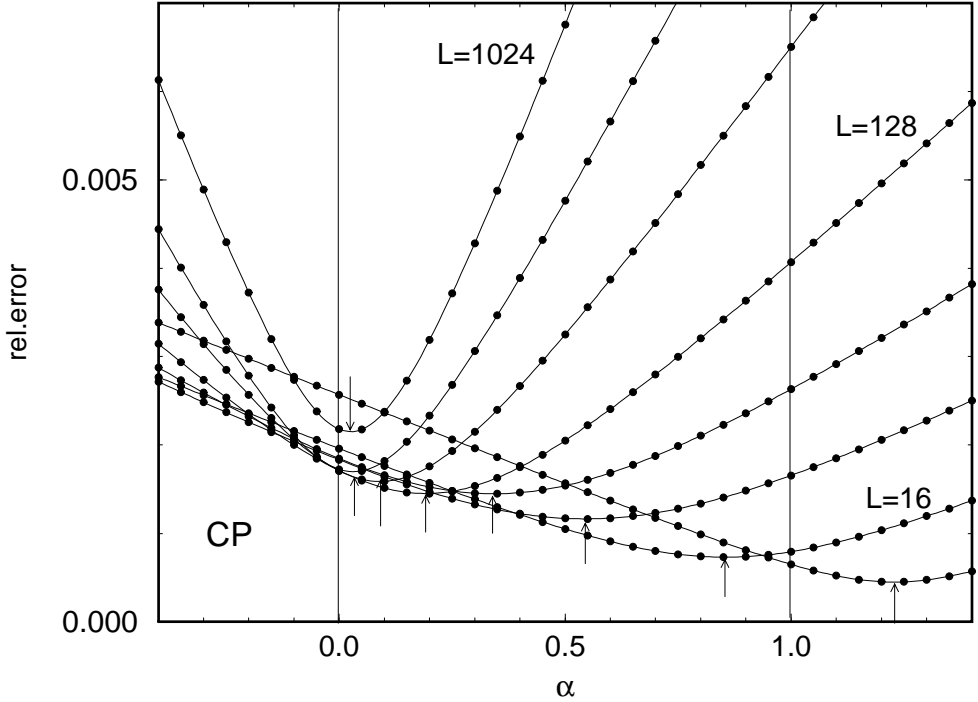}
\caption[Statistical error of the linearly combined energy estimator]{
    Relative statistical error 
    $\Delta \overline{U}_{\rm c}/\overline{U}_{\rm c}$ 
    of the linear combination 
    $\overline{U}_{\rm c}$ 
    of the two energy estimators for the convex potential as a 
    function of the interpolation parameter $\alpha$ (using the W-cycle
    update algorithm). Data symbols
    denote jackknife averages over $100$ blocks, and the solid lines 
    are computed according to the theoretical prediction 
    (\ref{eq:sigmacombinedgeneral}), using as input only the (jackknife) 
    variances 
    of the virial and the kinetic estimator (corresponding to 
    $\alpha=0$, resp. $\alpha=1$) and the (jackknife) covariance of
    the two estimators. The arrows indicate the values of optimal 
    $\alpha$ according to eq.~(\ref{eq:alphaopti}).
            }
\label{figure:interpol_cp}
\end{figure}

In order to illustrate the combination of the two estimators with
actual simulation data (using data for the convex potential obtained
by multigrid W-cycle updating), we show in
Fig.~\ref{figure:interpol_cp}
the relative statistical error
as a function of the interpolation parameter $\alpha$. 
In order to simulate the measurement of an arbitrary combination
of the two estimators,
the data points in Fig.~\ref{figure:interpol_cp} were obtained
from a times series of an estimator $U_{\rm c}(\alpha)$ for arbitrary 
$\alpha$ which was generated from the time series of each of
the individual measurements of $U_{\rm k}$ and $U_{\rm v}$.
The time series for $U_{\rm c}(\alpha)$ was then subsequently analysed for 
each $\alpha$ by the usual jackknife blocking procedure to obtain the error
estimate (cf.\ section~\ref{ssec:Block}).
The solid lines interpolating these data points, on the other hand,
were computed using 
eq.~(\ref{eq:sigmacombinedgeneral}) with the variances
and covariance of $U_{\rm k}$ and $U_{\rm v}$ obtained by the jackknife
method on the basis
of $100$ blocks. 

We see that the theoretically expected error
(\ref{eq:sigmacombinedgeneral})
reproduces indeed the empirically measured errors for a 
linear combination of the two estimators at each measurement.
We emphasize that the combination can hence always be done
post simulation
without any restriction, as long as the variances {\em and} the covariance
of the two estimators have been measured (i.e., it is {\em not\/}
necessary to store the complete time series of $U_{\rm k}$ and $U_{\rm v}$
which, in some applications, might cause disk-space problems). 

The arrows
indicate the locations of the optimal
values of $\alpha$ computed according to eq.~(\ref{eq:alphaopti}). 
These values
are also reported in Table \ref{table:cpukvc} together with the
correlation coefficient $\rho$ and the measured energies
using the optimal combination.  We note that for the coarsest
discretization of $L=8$ we do indeed find a case where the optimal
interpolation parameter $\alpha$ falls outside the usually expected
range of $[0,1]$.  Unfortunately, in this application the
theoretically possible error reduction for correlated estimators
is not realized, and the physically interesting data are those for large
$L$, where the correlation coefficient is only about $\rho = 0.3-0.5$.
Very similar plots were obtained for the case of the double-well
potential.

At this point it should be stressed that, in general, the question of how
much the error can be reduced in the simulation of a particular
problem at hand appears to be an empirical question.  As a general
lesson, we emphasize that the fact that two independent estimators are
correlated does not necessarily mean bad news.  The gain in efficiency
crucially depends on both the correlation and the ratio of the
individual variances.
A further discussion
will be given 
below in section \ref{ssec:Res_combi}.
%
       \section{Errors in the Monte Carlo process}          \label{sec:errors}
%
       In this paper we investigate both the variances and the
       autocorrelation times associated with different energy
       estimators.  While the variance is a property of the partition
       function and the estimator alone, the autocorrelation time also
       depends on the update scheme and represents the relevant
       quantitative measure for the dynamics of the Monte Carlo
       process.

In general, the autocorrelation time $\tau$ is proportional
to some power of the correlation length $\xi$ of the system,
$\tau \propto \xi^z$, with a dynamical critical exponent $z$.
For standard local update algorithms the exponent $z$ is close to
$2$ as can be argued in a simple random walk picture.
For spin systems or field theoretic models undergoing a second-order
phase transition the correlation length diverges
and as soon as $\xi$ exceeds the linear size $L$ of the lattice the
autocorrelation times diverge like $\tau \propto L^z$.
In statistical mechanics and lattice field theory this problem
is known under the name of {\em critical slowing down\/}.

In path-integral simulations a very similar problem occurs in the continuum
limit $\eps\rightarrow 0$ with $\beta$ fixed, or, equivalently,
$L\rightarrow \infty$. The reason for this slowing down is easily
understood. The correlations $\la x_k x_{k+l}\ra$ only depend on
$\beta$ and on the gaps between the energy levels. Hence the
correlation length $\xi$ only depends on the physical parameters
at hand, and consequently always diverges linearly with $L$ if measured
in units of the lattice spacing $\eps$. Thus we expect that
the autocorrelation time for path-integral
simulations grows as
\begin{equation}
\tau \propto L^z,
\label{eq:tau}
\end{equation}
with a dynamical critical exponent $z$ as well, and that for standard local
algorithms $z=2$.
Note that in contrast to the infinite-volume limit in statistical
mechanics, for path integrals also the Hamiltonian, i.e., the exponent in
(\ref{eq:Z_PI_dis}) changes its form
in the limit $L\rightarrow \infty$ which is a characteristic
feature of path-integral simulations.
As a consequence slowing down
occurs in the continuum limit for {\em any\/} fixed $\beta$ and
{\em any\/} set of potential parameters.

\subsection{Definition and measurement of autocorrelation times}\label{ssec:Tau}

Since the measurement of autocorrelation times will be of central
concern in the remainder of this paper let us briefly recall the
basic definitions \cite{sokal92,ms88}.

In general, if ${\cal O}_i$ denotes the $i$th measurement of an
observable ${\cal O}$ in the Monte Carlo process
the (normalized) autocorrelation function $A(j)$ is defined by
\begin{equation}
      A(j) = \frac{\langle{\cal O}_i{\cal O}_{i+j}\rangle
                 - \langle{\cal O}_i\rangle ^2}
                  {\langle{\cal O}_i^2\rangle
                 - \langle{\cal O}_i\rangle ^2}.
 \label{eq:A(j)}
\end{equation}
In Monte Carlo simulations we are always dealing with a {\em finite}
number of measurements $N_m$. Nevertheless, computing the variance for
the mean $\overline{{\cal O}}$ of measurements is straightforward and
yields an error estimate of the form $\Delta \overline{{\cal O}} =
\sqrt{2 \tint(N_m)} \sqrt{\sigma^2/N_m}$, where $N_m$ is the number of
measurements used to compute the mean value $\overline{{\cal O}}$.
Here $\tint(k)$ is an integrated autocorrelation time given by~\cite{jan96}
\begin{equation}
      \tint(k) = \frac{1}{2} + \sum_{j=1}^k A(j)
                \left[1-\frac{j}{N_m}\right].
 \label{eq:tau(k)}
\end{equation}
The effective statistics is thus reduced to $N_{\rm eff} =
N_m/2\tint(N_m)$.  Or, in other words, to achieve a given error $\Delta
\overline{{\cal O}}$ the run-time (i.e. the budget) has to be
increased by a factor of $2\tint(N_m)$.

For large $j$ the autocorrelation function $A(j)$ usually decays like
an exponential
\begin{equation}
      A(j) \buildrel j \rightarrow \infty \over \longrightarrow
            a \exp(-j/\texp),
 \label{eq:Afit}
\end{equation}
where $\texp$ denotes the exponential autocorrelation time and $a$ is
some constant.
Since, in general, all these quantities depend on the observable under
consideration we will indicate the relevant observable by an additional
subscript unless it is clear from the context which observable is meant.

Due to the usual exponential decay of the autocorrelation function
$A(j)$ and the large number of measurements $N_m$ in a Monte Carlo
simulation, the factor $\left[1-\frac{j}{N_m}\right]$ in
(\ref{eq:tau(k)}) accounting for the finite size of the statistical
sample may in practical applications savely be neglected. In
practice, one therefore usually computes the autocorrelation time by
computing the sum in eq.~(\ref{eq:tau(k)}) without the correction
factor and by cutting it off self-consistently at some $k_{\rm
max}$ such that $n_{\rm cut}\tint(k_{\rm max})\approx k_{\rm
max} << N_m$. Usually $n_{\rm cut}$ is set equal to $n_{\rm cut}=6$ or $8$.
As long as integrated and exponential autocorrelation times are
roughly the same this method gives reliable estimates for $\tint$. If
exponential and integrated autocorrelation times are very much
different from each other this method tends to underestimate the
integrated autocorrelation time.
This typically happens when the faster decaying modes neglected 
in (\ref{eq:Afit}) are still important for relatively large time lags $j$.

A more reliable way of determining the autocorrelation times in this case
is to rewrite the integrated autocorrelation time as~\cite{js95}
\vspace*{-0.2cm}
\begin{eqnarray}
      \tint(k) & = & \tint -
                     a \sum_{j=k+1}^{\infty} \exp(-j/\texp)
        \\
           & = & \tint - a \frac{\exp\{-1/\texp\}}
               {1-\exp\{ -1/\texp \}}
                \exp\{-k/\texp\},
           \label{eq:Bfit}
\end{eqnarray}
where $\tint \equiv \tint(\infty)$. A three-parameter fit of $\tint(k)$ then
gives both the integrated and the exponential autocorrelation times $\tint$
and $\texp$ simultaneously, which is a further advantage of this method.

\begin{figure}[tbh]
\vskip 6.0truecm
\includegraphics{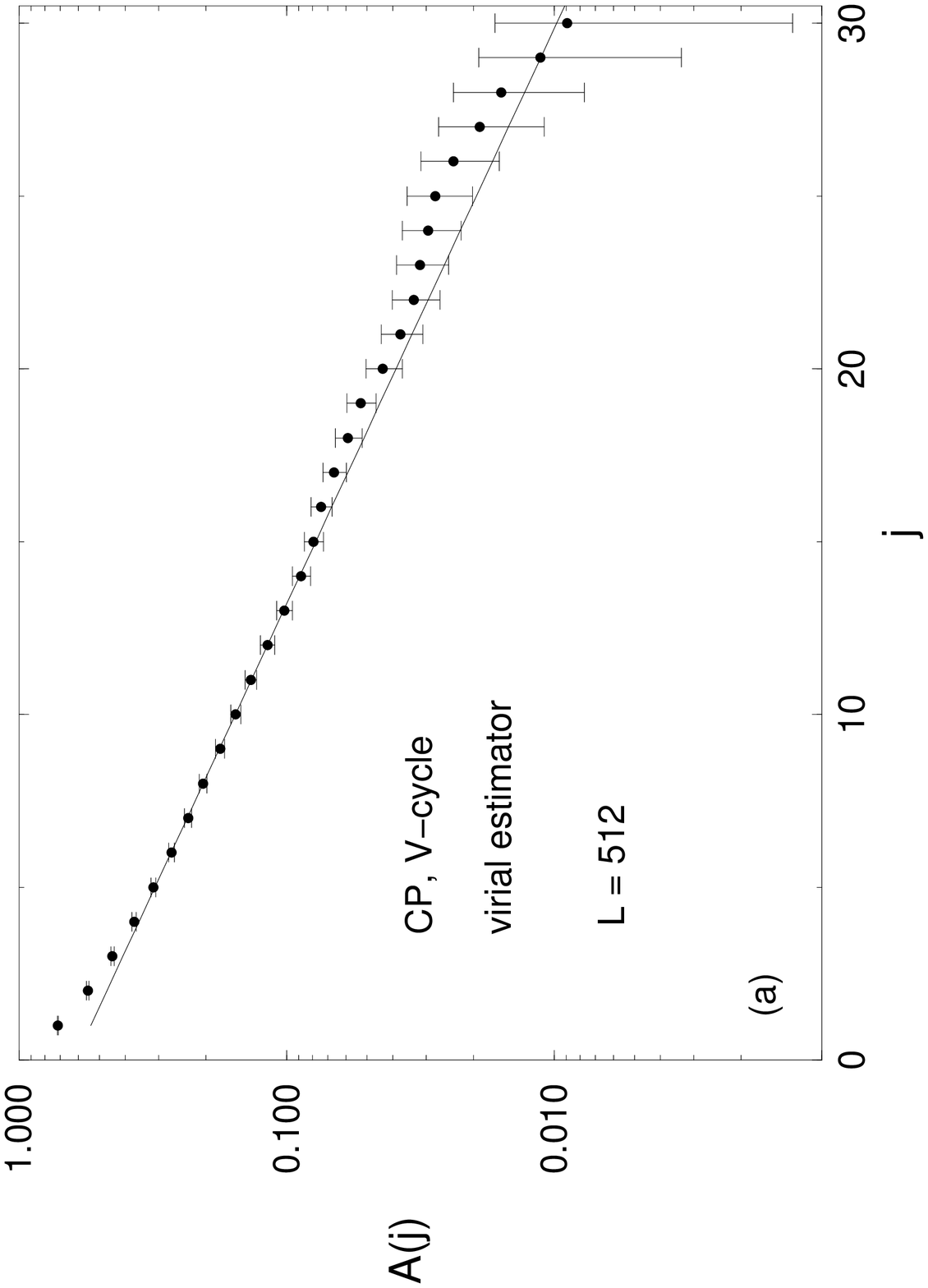}
\vskip 7.7truecm
\includegraphics{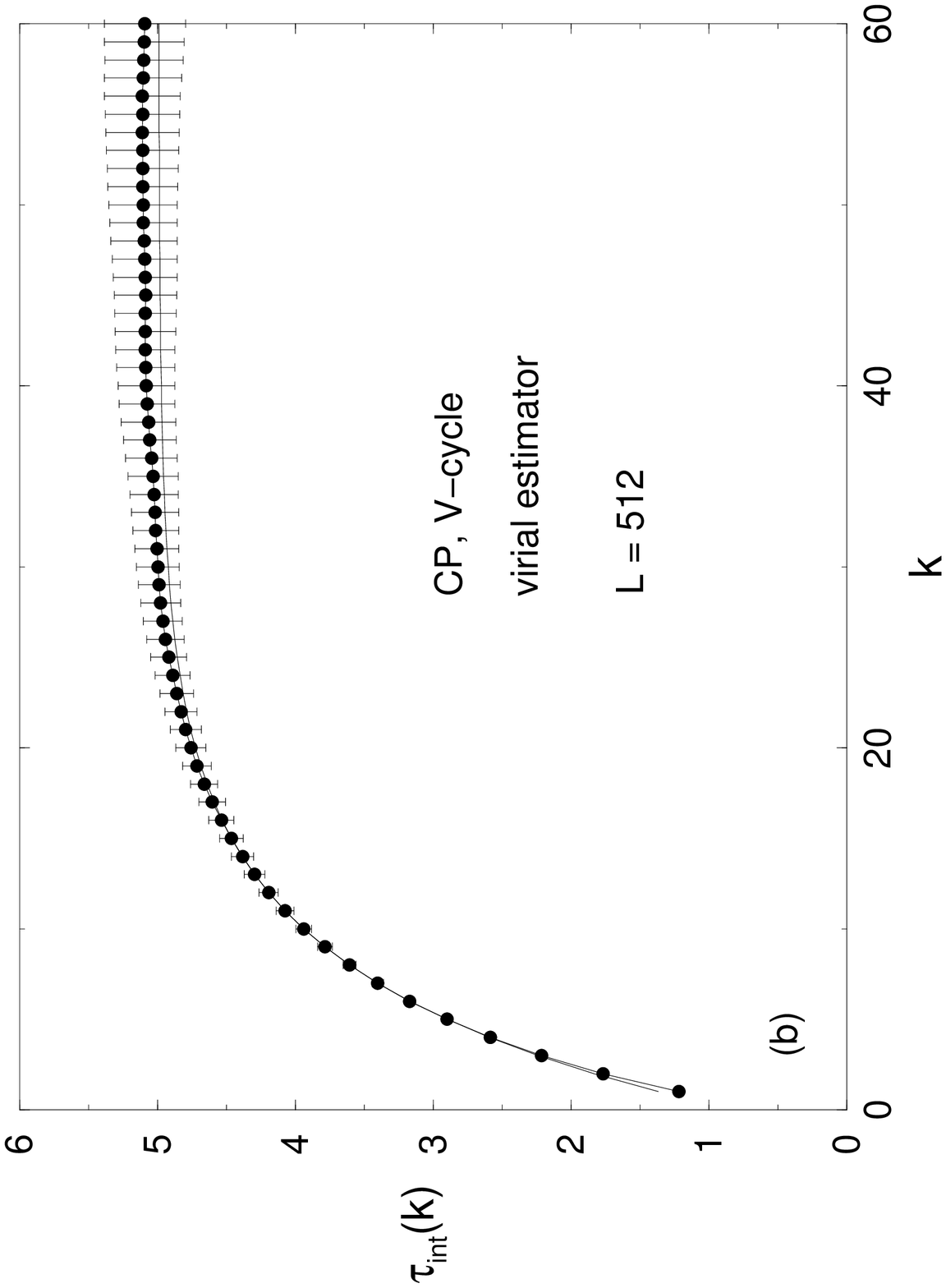}
\caption[a]{(a) The autocorrelation function $A(j)$ on a logarithmic scale
and (b) the integrated autocorrelation time $\tint(k)$ of the virial
estimator as obtained with the V-cycle multigrid algorithm for the
convex potential (CP) at $\beta=10$ and $L=512$.  Solid lines show
fits according to eq.~(\ref{eq:Afit}) resp. (\ref{eq:Bfit}). The
asymptotic value of $\tint(k)$ quoted in Table~\ref{table:cptau} is
$\tau_{\rm int,v} = 5.07(21)$.}
\label{figure:autocorr}
\end{figure}

In Fig.~\ref{figure:autocorr} the behaviour of $A(j)$ and $\tint(k)$
of the virial estimator is illustrated for one typical example (CP
potential, V-cycle multigrid update, $\beta=10$, $L=512$). A two-parameter 
fit of
the autocorrelation function $A(j)$ according to eq.~(\ref{eq:Afit})
in the range $j=5,\dots,15$ for this case gave values of $a=0.619(33)$
and $\tau_{\rm exp,v} = 7.23(40)$. A three-parameter fit of the
integrated autocorrelation time according to eq.~(\ref{eq:Bfit}) in the
same range yielded values of $a=0.614(42)$, $\tau_{\rm exp,v}=7.26(48)$,
and $\tau_{\rm int,v}=4.98(15)$.  A determination of the integrated
autocorrelation time via a self-consistent cutoff at $8\tau_{\rm int,v}$ 
on the
other hand yielded a value of $\tau_{\rm int,v} = 5.07(21)$
(cp.\ Table~\ref{table:cptau}). Clearly all values are consistent
within error bars.
\vspace*{-0.2cm}
\subsection{Blocking and jackknife procedures}    \label{ssec:Block}

Another way of estimating the true error of the Monte Carlo simulation
is to divide the $N_m$ measurements into $n_{\rm bl}$ blocks
of size $N_{\rm bl} = N_m/n_{\rm bl}$ and compute the averages for each block
separately. The block averages are then again stochastic variables
with the same mean but reduced autocorrelations. In fact, if the
block length is appreciably larger than the integrated autocorrelation
time for the observable under consideration ($20\tint$, say, would
clearly be sufficient) than the block averages are (almost)
uncorrelated and we can estimate the error of the observable
from the variance of the block averages,
$\Delta\overline{\Obs}=\sqrt{\sigma^2_{\Obs,{\rm bl}}/n_{\rm bl}}$.
Since $n_{\rm bl}$ should be at least around 20 to allow for a statistically
meaningful estimation of $\sigma^2_{\Obs,{\rm bl}}$, the number of
measurements per block is always much smaller than the total number
of measurements, $N_{\rm bl} \ll N_m$. For observables that can be estimated
only with so-called biased estimators this may result in severe
systematic errors when $N_{\rm bl}$ becomes too small.

It is therefore gratifying that bias-reduced and thus more accurate
estimates of the errors can be achieved by using so-called jackknife
blocking techniques~\cite{jack}. The main difference is that here the
block averages are taken over the whole run with only one block
excluded. These jackknife block averages have consequently a much
reduced variance $\sigma^2_{\Obs,{\rm j-bl}}$ and the error estimate
now reads $\Delta\overline{\Obs}=\sqrt{(n_{\rm
    bl}-1)\sigma^2_{\Obs,{\rm j-bl}} /n_{\rm bl}}$, where the factor
$(n_{\rm bl} - 1)$ corrects for the trivial correlations between
different jackknife blocks (because each measurement enters in {\em
  all but one\/} jackknife block -- this has nothing to do with
autocorrelations). It should be stressed that for unbiased estimators,
such as arithmetic mean values, the blocking and jackknife blocking
method give identical results. The advantages of jackknife blocking
only show up for biased estimators, such as those for the specific
heat or autocorrelation functions.

Using the blocking or jackknife blocking error estimate the
integrated autocorrelation time for an observable $\Obs$
can then also be obtained by
inverting the standard error formula as
\begin{equation}
        \tint = \frac{(\Delta\overline{\Obs})^2N_m}{2\sigma_{\Obs_i}^2},
\end{equation}
where $\Delta\overline{\Obs}$ is the error measured in the blocking analysis
and $\sigma_{\Obs_i}^2$ is
the variance of the single (unblocked) measurements. It should be
stressed that this method of estimating the integrated autocorrelation
time is only valid if the block length $N_{\rm bl}$ is several times
larger than the true integrated autocorrelation time. Otherwise
$\tau_{\rm int}$ will be systematically underestimated. At the same time we
want to have, of course, as many blocks as possible in order to reduce the
statistical error on the error estimates. Since $n_{\rm bl} N_{\rm bl} = N_m$
is fixed for a given statistics, these are conflicting requirements. If the
measurements are not too strongly correlated, a reasonable compromise is the
choice $n_{\rm bl} \approx N_{\rm bl} \approx \sqrt{N_m}$.
\clearpage
%
            \section{Update algorithms}                     \label{sec:updates} 
%
Before presenting our results in the 
next section, 
we will briefly
review the update algorithms we have investigated, notably the 
multigrid method and the staging algorithm.

\subsection{Local algorithms}                     \label{ssec:Local}

The most extensively used algorithm for PIMC simulations
still is the standard Metropolis algorithm~\cite{metro}. As far as
accuracy and efficiency are concerned it has the well-known serious
drawback of {\em local\/} algorithms that in the continuum
limit $\epsilon = \beta/L \longrightarrow 0$
the autocorrelation times diverge 
quadratically with the grid size, $\tau \propto L^2$,
resulting in a severe slowing down of the Monte Carlo process, see the
discussion in section~\ref{sec:errors}. In spite of this drawback it
is still widely used for its simplicity even though a lot of computer
time might be saved by using more refined algorithms.

A local algorithm which does reduce slowing down to some extent
is the hybrid overrelaxation method~\cite{over}.
This algorithm mixes
(deterministic) overrelaxed updates of the path with stochastic
Metropolis sweeps. If the ratio of overrelaxed and Metropolis sweeps
is chosen properly, i.e., in proportion to the spatial correlation
length, this algorithm reduces slowing down to a linear divergence~\cite{js92b,thesis}, $\tau \propto L$.

More successful are {\em non-local\/} algorithms which will
be briefly described in the next two subsections.

\subsection{Multigrid method}                     \label{ssec:Multi}

The basic idea of the multigrid approach~\cite{multigrid,hac85} is to
perform non-local updates of the variables by working on a set of
successively coarser discretizations of the time axis (``grids'') in
order to take into account long wavelength fluctuations of the paths
more efficiently.
To this end one performs collective updates on different length scales
by visiting various coarsened grids in a systematic order as
extensively discussed in the context of partial differential
equations~\cite{hac85}. The auxiliary variables on the coarsened grids 
are then
interpolated back to the finer grids and eventually to the original
grid using some specific interpolation scheme in a recursive
manner. Equivalently, one may also view the multigrid approach from a
unigrid point of view where the update on a coarsened grid corresponds
to a simultaneous move of a group of adjacent variables on the
original grid. Using the so-called piecewise constant
interpolation scheme for example, this amounts, in the unigrid
viewpoint, to proposing moves for blocks of 
$1, 2^d, 4^d,\dots, V=L^d=2^{nd}$ 
adjacent variables in conjunction. Here $d$ is the
dimension of the system under consideration. For PIMC
simulations we usually have $d=1$, even for quantum chains or quantum crystals
since it may well suffice to use multigrid acceleration only along the
Trotter direction of the discretized path integral \cite{js95b}. 
Particular
successful sequences of length scales $2^k$ are the so-called V-cycle
with $k=0,1,\dots,n-1,n,n-1,\dots,1,0$, and the W-cycle whose
graphical representation looks like the letter W (for $n=3$, e.g.,
this is k = 0, 1, 2, 3, 2, 3, 2, 1, 2, 3, 2, 3, 2, 1, 0) \cite{athens}.
The update at level $k$ thus consists in considering a common move
$\Delta x$ for all $2^{kd}$ variables of one block, $x_k
\longrightarrow x_k + \Delta x$, $i \in {\rm block}$, computing the
associated energy change and applying the usual accept/reject
criterion. 

The multigrid approach thus has a number of parameters
which may be adjusted to suit the problem at hand, notably the choice
of the interpolation scheme, the block length, the recursion scheme
(e.g. V- or W-cycle), the number of updates on each grid in going down
the recursion scheme (presweeps) and in going up again (postsweeps),
and the Metropolis parameters for the coarsened grid updates. 
In the Gaussian case it is known that by using the piecewise constant 
interpolation scheme the V-cycle leads to a linear divergence of 
autocorrelation times, $\tau \propto L$, while the more successful W-cycle
beats slowing down completely, $\tau \propto {\rm const}$ \cite{js93a}.

Let us finally emphasize that the multigrid method can easily be applied 
to general $d$-dimensional lattice field theories. In fact, it is in this 
context where the stochastic multigrid Monte Carlo formulation appeared 
first in the literature~\cite{multigrid}, and only quite recently these 
ideas have been adapted to path-integral simulations \cite{js92b,js93a,js94}. 

\subsection{Staging algorithm}                  \label{ssec:Staging}

The basic idea of the staging algorithm~\cite{staging} is to rewrite
the discretized quantum statistical partition function
(\ref{eq:Z_PI_dis}) in such a way that a sequence of $j$ adjacent
variables can be updated one by one independently, i.e., by effectively
removing the coupling between the variables.  The coupling between
variables stems from the kinetic term and, in the continuum limit,
$\epsilon\rightarrow 0$, the kinetic term
$m/(2\epsilon)(x_k-x_{k-1})^2$ numerically starts to dominate over the
potential energy term $\epsilon V(x_k)$. The kinetic term for adjacent
variables can, however, trivially be rewritten as
\begin{equation}
 (x_{k+1}-x_{k})^2 +
 (x_{k+2}-x_{k+1})^2=
 (x_{k+2}-x_{k})^2/2 +
 2(x_{k+1}-x_{k+1}^{\ast})^2,
\end{equation}
with $x_{k+1}^{\ast}\equiv (x_{k}+x_{k+2})/2$. The variable
$x_{k+1}$ can thus be decoupled from its neighbours. The crucial
observation for the staging algorithm is that this can be iterated as
\begin{equation}
 \frac{1}{j}(x_{k+j}-x_{k})^2 +
 (x_{k+j+1}-x_{k+j})^2 =
 \frac{1}{(j+1)}(x_{k+j+1}-x_{k})^2 +
 \frac{j+1}{j}(x_{k+j}-x_{k+j}^{\ast})^2,
\end{equation}
with new variables $x_{k+j}^{\ast}\equiv(x_{k}+x_{k+j+1})/(j+1)$.
A sequence of free particle propagators can thus be rewritten as~\cite{staging}
\begin{eqnarray}
 &&\prod_{i=0}^j \langle x_{k+i} | \exp(-\epsilon\frac{1}{2m}\hat{p}^2) |
                       x_{k+i+1} \rangle = \\
 &&\langle  x_{k} | \exp(-j\epsilon\frac{1}{2m}\hat{p}^2) |
                       x_{k+j+1} \rangle
 \prod_{i=2}^{j}\left[\frac{m_i}{2\pi\epsilon}\right]^{1/2}
          \exp\left(-\frac{m_i}{2\epsilon}(x_i-x_i^{\ast})^2\right),
     \nonumber
\end{eqnarray}
where renormalized masses $m_i\equiv\frac{i}{i-1}m$ have been
introduced. Selecting the end points $x_{k}$ and $x_{k+j+1}$ of some
segment of the discretized path with $j$ ``beads'' in between, one can
thus perform a (recursive) change of variables $x_{k+i}\rightarrow
x_{k+i}^{\ast}$, $i=1,\dots j$, in the discretized partition function
(\ref{eq:Z_PI_dis}). For the staging segment, this would eliminate the
nearest neighbour coupling stemming from the kinetic energy. For the
variables of the staging segment the partition function hence reduces
to a collection of independent oscillators moving in an external
potential which depends on the transformation of the variables. The
staging variables may then be updated using Gaussian random variables
$u_{k}\equiv x_{k}-x_{k}^{\ast}$ with ``masses'' $m_i$, and a
Metropolis like acceptance rule for the external potential.

In contrast to the multigrid method the staging algorithm only allows
for one single tunable parameter, the length $j$ of the staging
segment. The optimal choice of the staging parameter $j$ depends on
the observable of interest, its estimator and on the discretization.
But since it is the greater mobility of the variables of the staging
segment which reduces autocorrelation times the optimal staging length
$j_{\rm opt}$ scales with the correlation length along the path, i.e.,
it scales with the number of variables $L$ in the discretized
partition function~\cite{js96a}. For an appropriate choice of $j_{\rm
  opt}$ the staging algorithm then completely reduces slowing down in
the continuum limit, $\tau \propto {\rm const}$.  Thus, both the
staging and multigrid W-cycle represent refined PIMC update schemes
which beat the continuum slowing down. Elsewhere we have compared the
two algorithm and discussed their mutual merit in more
detail \cite{js96a}.
%
                  \section{Results}                        \label{sec:results}
%
\subsection{Simulation details}                    \label{ssec:Simu}

For each of the four update algorithms described above (Metropolis, multigrid
V- and W-cycle, staging algorithm)
we simulated the path integral (\ref{eq:Z_PI_dis}) for grids of 
size $L = 2^3 = 8$ up to $L = 2^{10} = 1024$ sites.
In all our simulations the mass $m$ was set equal to $1$ and
the inverse temperature was equal to $\beta = 10$.
In our simulations we performed $N_m=100\,000$ updates of the path
after discarding $5\,000$ initial updates of the path for thermalization.

In the case of the Metropolis algorithm an ``update of the path'' here
means $n_e$ sweeps over the full path with single-hit updates of each
site with roughly $50\%$ acceptance probability. In order to
accurately assess autocorrelation times, the parameter $n_e$ was
adjusted in such a way that the autocorrelation times in units of
measurements were comparable for all grid sizes $L$. For the convex
potential this could always be achieved by setting $n_e = 1$, except
for $L=256$, $512$, and $1024$ where we had $n_e=5$, $20$, and $80$,
respectively.
For the double-well potential we started out with $n_e = 1$, 1, 1, and
2 for $L=8$, 16, 32, and 64. For the larger grid sizes, however, the
autocorrelation times in units of single sweeps turned out to be so
different for the two estimators that we actually performed two sets
of simulations with different choices of $n_e$ (the combined estimator
then obviously makes no sense). In the first set, focusing on the
autocorrelation time for the kinetic estimator, we could still use
$n_e=1$, except for the largest grid $L=1024$ where we set $n_e=6$. In
these runs the autocorrelation times for the virial estimator turned
out to be far too big to be measurable on the larger grids. To satisfy
our own curiosity and to measure autocorrelation times for the virial
estimator as well, we therefore performed a second set of simulations
with $n_e = 15$, $60$, $150$, and $600$ for $L=128$, $256$, $512$, and
$1024$, respectively, even though this clearly requires employing CPU
resources out of proportion. In the following, all autocorrelation
times for the Metropolis algorithm will be given in units of single
sweeps.

In the case of the multigrid algorithm an ``update of
the path'' means a complete V- resp.  W-cycle with $n_{\rm pre}=1$
presweeps and $n_{\rm post}=0$ postsweeps. On each grid we performed
single-hit Metropolis updates with an acceptance rate of $40\%-60\%$
which for the system at hand could be achieved with the same maximal
step width on all grids.  

In the case of the staging algorithm, an
``update of the path'' means ${\rm int}(L/(j-1))$ calls to the staging
routine which moves $j-1$ adjacent variables at each call.  The choice
of the staging length $j$ is shown in Table~\ref{table:jopt} below. For the
analysis of the 
combined estimator we used the same $j$ as for the virial estimator. This
will be discussed in more detail below in section \ref{ssec:Res_combi}.
Notice that the above definition 
in general implies updates of less than $L$
variables. We have therefore rescaled the actually measured
autocorrelation times by a factor $({\rm int}(L/(j-1)))/(L/(j-1))$.
This enables a direct comparison of the staging algorithm with the
standard Metropolis algorithm. A comparison with the multigrid method
has to take into account a constant factor for the V-cycle (which of 
course depends on the implementation but should theoretically be $\approx 2$)
and an extra factor of $\log L$ for the W-cycle.

After each update of the path we measured the internal energy using
both the ``kinetic'' estimator (\ref{eq:kineticestimator}) and the
``virial'' estimator (\ref{eq:virialestimator}).  The time series of
these measurements were then analysed by jackkniving the data on the
basis of $100$ blocks. Autocorrelation times were obtained by
cutting $\tau_{\rm int}(k)$ self-consistently at $k_{\rm max}=8$.
The reported errors for the
autocorrelation times were again obtained by jackkniving. 

With these remarks in mind we emphasize that our data now allow for a
precise and detailed comparison of the commonly used energy estimators
taking into account the dynamics of different update schemes.

\subsection{Results for the energy}              \label{ssec:Energy}

Tables \ref{table:cpukvc} and \ref{table:dwukvc} show the measured
energies for the convex potential (CP) and for the double-well potential (DW)
using the ``kinetic'' estimator
and the ``virial'' estimator as well as using the optimally combined
estimator. For ease of comparison we also 
give for each energy
measurement the relative error in percent. In addition, for the combined
estimator, we list the correlation coefficient $\rho$
and the parameter $\alpha_{\rm opt}$ used for computing
the combined estimator according to eq.~(\ref{eq:linearcombi}).

Since in our simulations we saved the time series of the $U_{\rm k}$-
and $U_{\rm v}$-measurements, we were able to
compute a posteriori
time series of $U_{\rm c}$ for any $\alpha$ which could then be
analyzed in the same fashion as the (run-)time series for the
``kinetic'' and the ``virial'' estimator. The energy values and the
jackknife errors reported in Tables \ref{table:cpukvc} and
\ref{table:dwukvc} were thus obtained by analyzing the time series of
the combined estimator $U_{\rm c}(\alpha_{\rm opt})$ for the optimal
choice of $\alpha$.  From the discussion in section \ref{ssec:Combi},
in particular Fig.~\ref{figure:interpol_cp}, it should be clear that
we could equally well have applied error propagation using the
variances and the covariance of the two original estimators.

The analysis of a times series of $U_{\rm c}$ has the advantage that,
for a more detailed discussion, the variance of the individual
measurement and the autocorrelation time of the combined estimator can
be computed as well.  In Tables \ref{table:cptau} and
\ref{table:dwtau} we list for each estimator the variance $\sigma_{\rm
k,v,c}^2$ of the individual measurements as well as their integrated
autocorrelation times $\tau_{\rm int,k,v,c}$.  As already mentioned,
the reported errors for these quantities were obtained using
jackknife blocking.

Looking at the energy values in Tables \ref{table:cpukvc} and
\ref{table:dwukvc}, the first thing we notice is that all three
estimators give indeed compatible energy estimates within error bounds
and all estimates converge to the correct continuum energy of
$U=0.80377$ (CP) resp. $-0.903965$ (DW).
These values can be easily obtained by numerical integration
of the associated Schr\"odinger equation. For the convex potential
this is basically the ground-state energy $E_0$, since already the first
excited state with 
$E_1 = 2.736$ is strongly
suppressed at $\beta=10$ and does not affect the 
significant digits of $U$.
For the double-well potential
the value of $U$ was obtained by using
$E_0 = -0.913371$, $E_1 = -0.892348$, $E_2 = 0.029846$, and $E_3 = 0.37813$.

One also recognizes a distinct finite-size dependence which on theoretical
grounds should be proportional to $L^{-2}$. For the most accurate
measurements in our simulations (using the combined estimator and the
W-cycle) we find that only for $L\geq 256$ (CP) resp. $L\geq 128$ (DW)
the discretization error is no longer relevant.
We mention in passing that this systematic error may be reduced
using the Takahashi-Imada scheme~\cite{ti84} based on higher-order Trotter
decomposition. Since this modification only amounts to
adding a local term to the potential energy this scheme
is perfectly compatible with all update algorithms discussed here, and as 
far as the 
qualitative behaviour of
statistical errors is concerned we do not expect any significant
deviations from the conclusions drawn in this study.

\subsection{Autocorrelation times and statistical efficiency} \label{ssec:Effi}

\begin{figure}[bhp]
\vskip  2.0truecm
\includegraphics{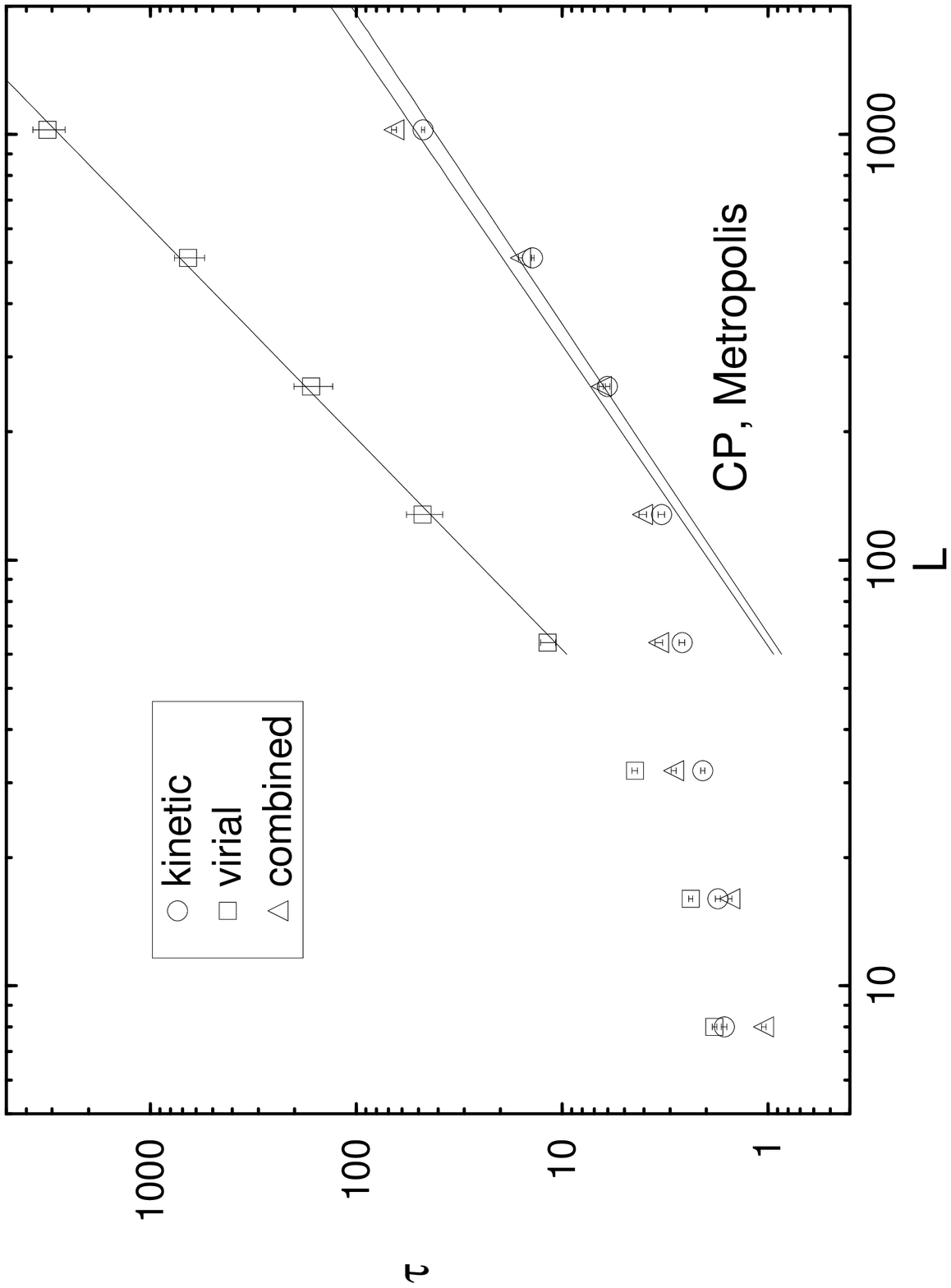}
\includegraphics{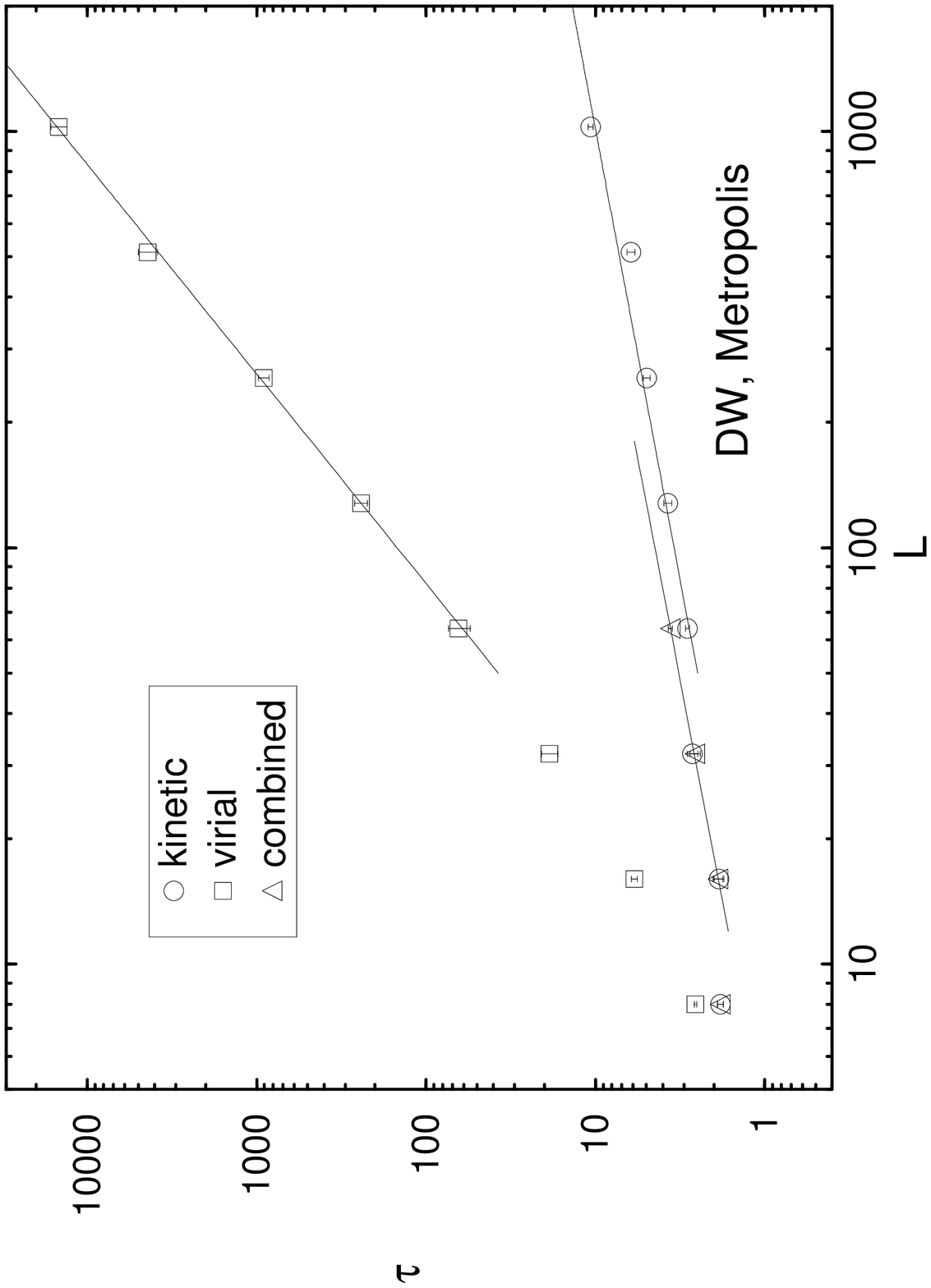}
\vskip 5.2truecm
\includegraphics{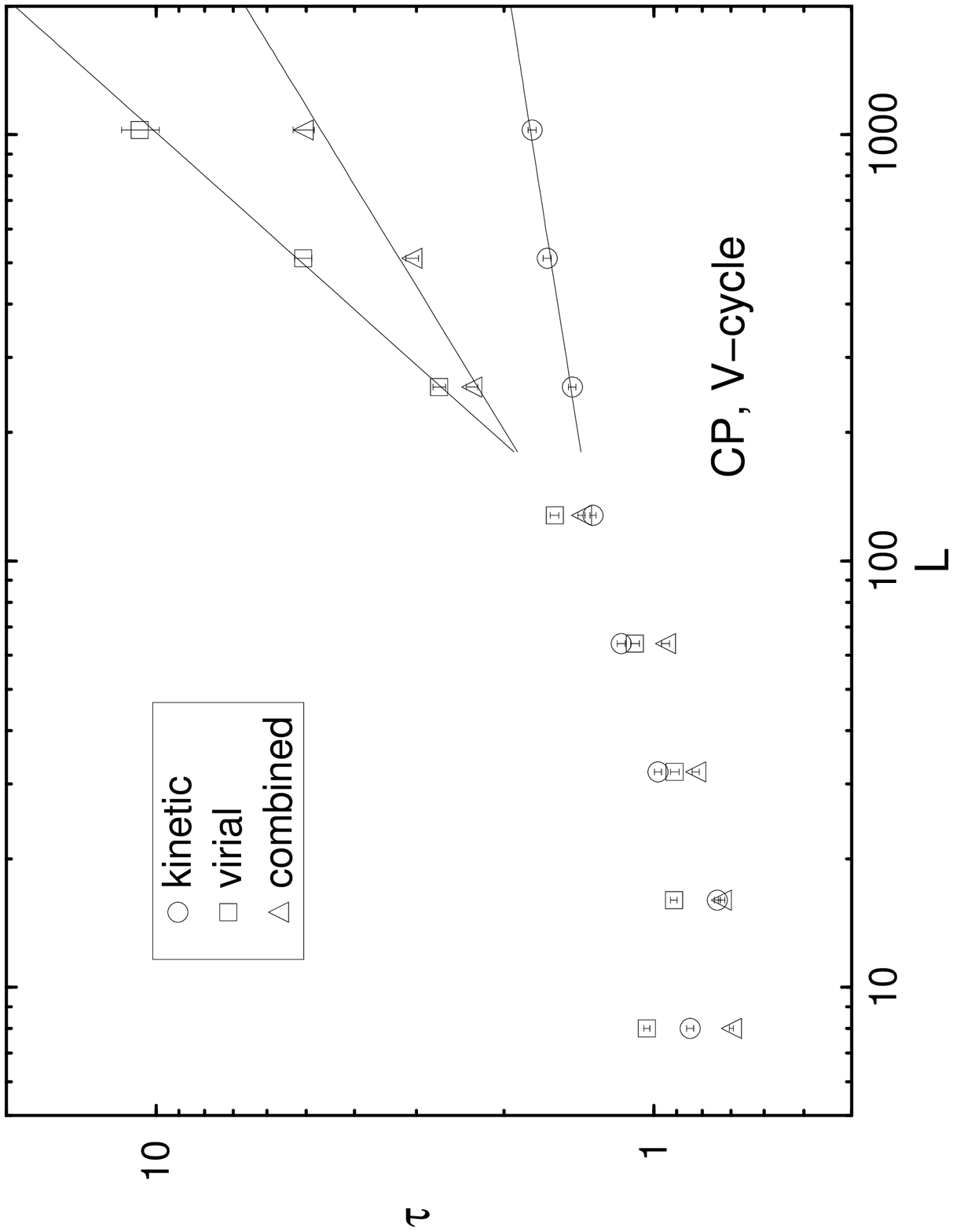}
\includegraphics{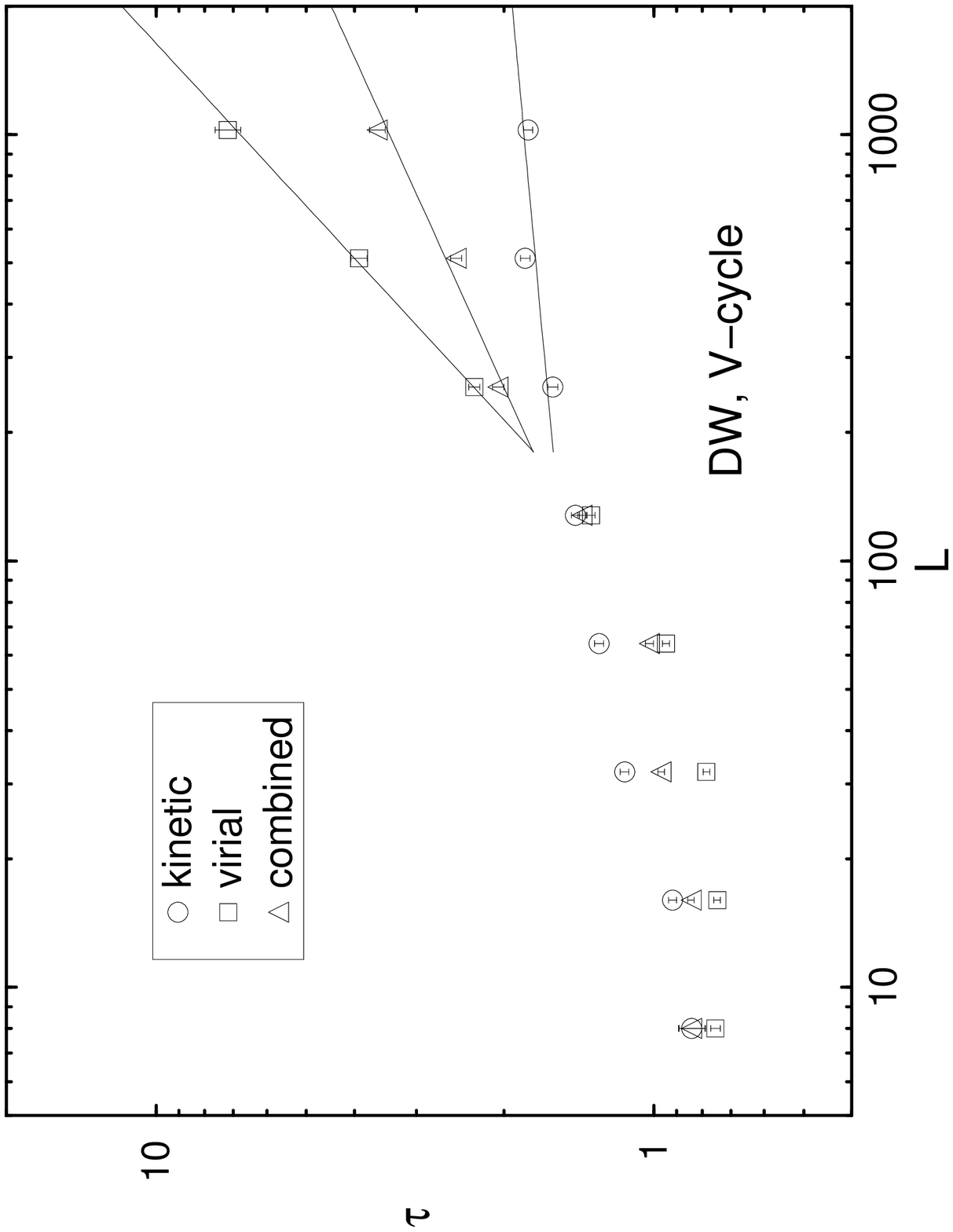}
\vskip 5.2truecm
\includegraphics{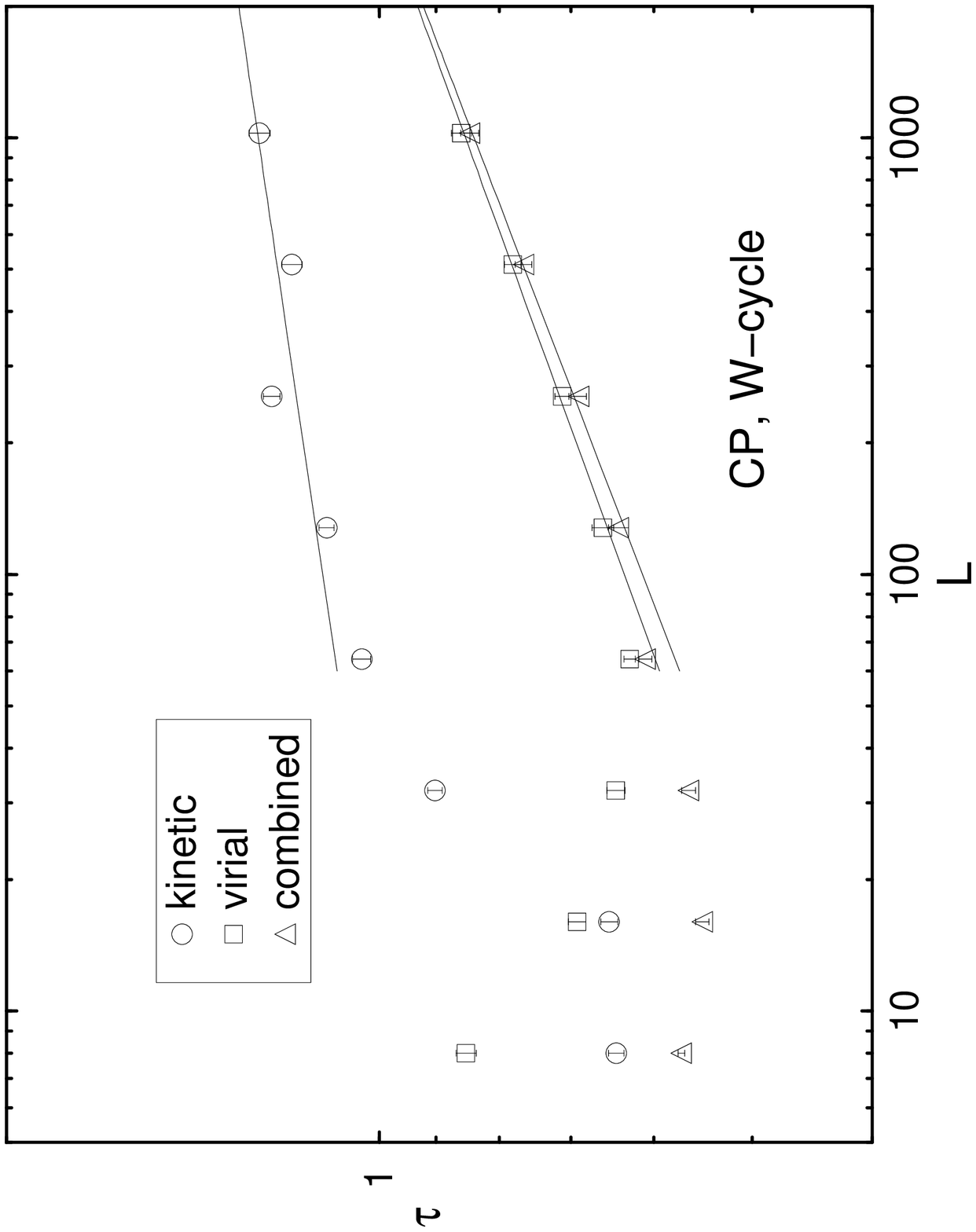}
\includegraphics{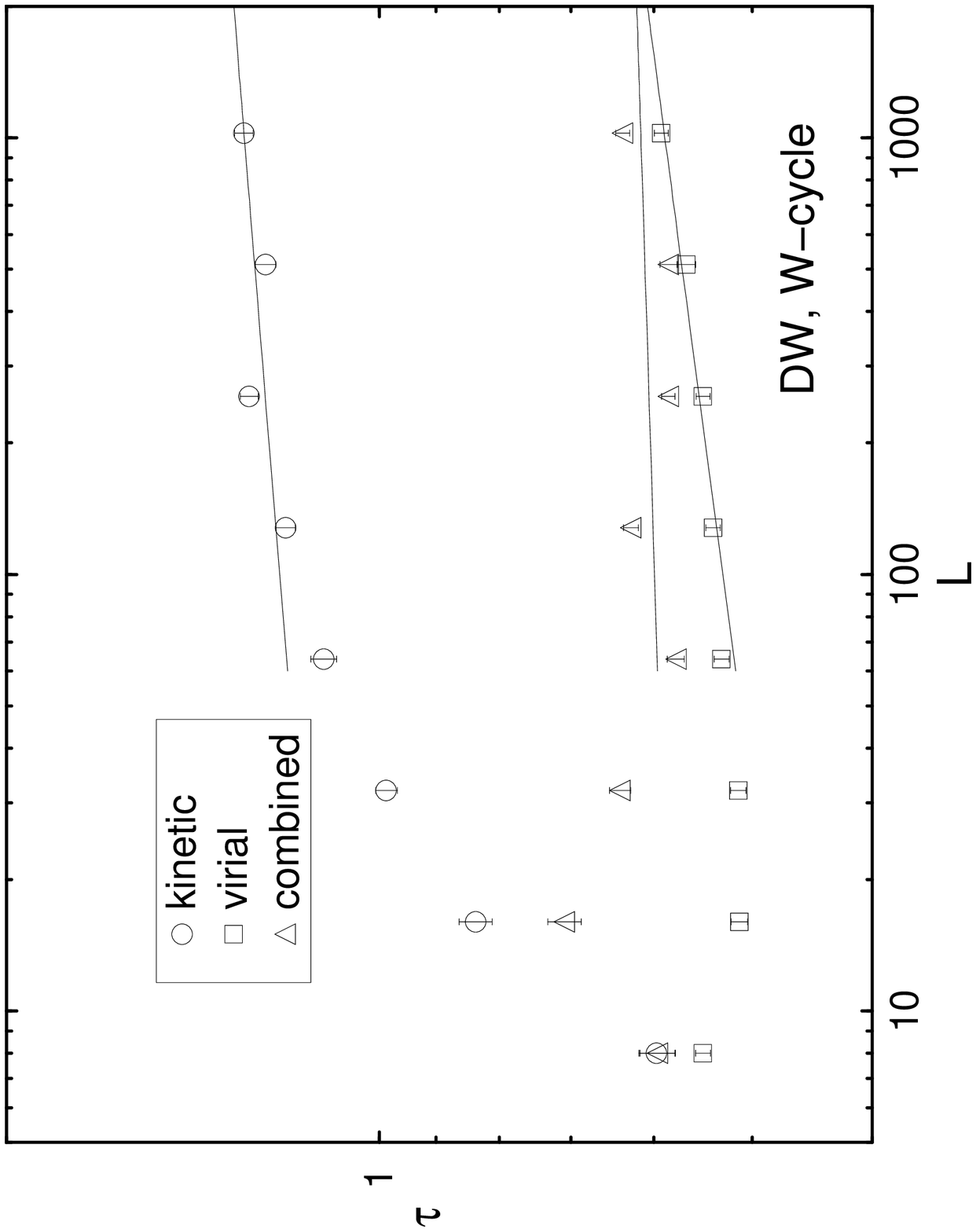}
\vskip 5.2truecm
\includegraphics{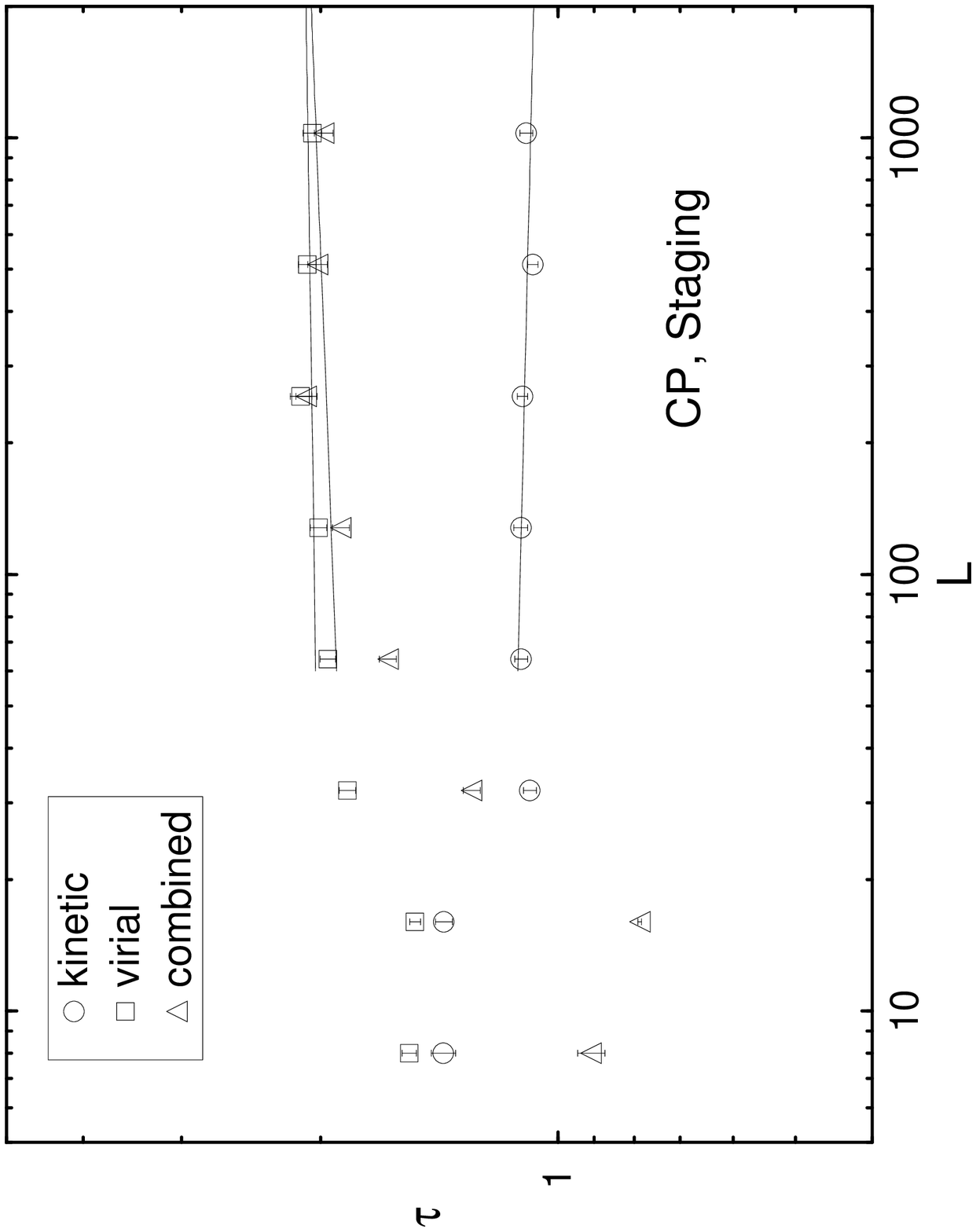}
\includegraphics{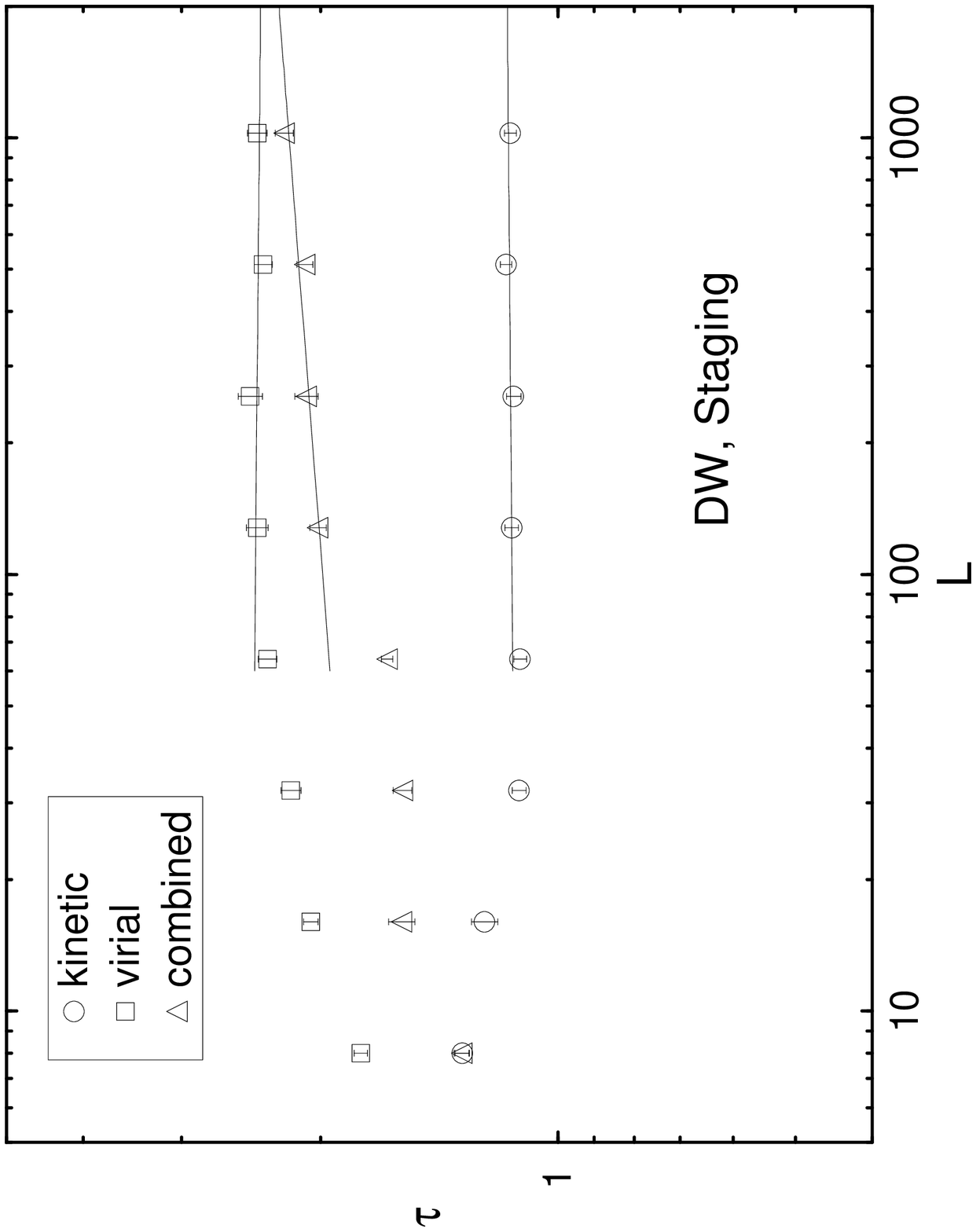}
\caption[a]{{
Integrated autocorrelation times $\tau_{\rm int}$ on a logarithmic scale 
for the three energy estimators
using different update algorithms for the convex potential (CP)
and the double well (DW).
Straight lines show fits of the form $\tau_{\rm int} = \alpha L^z$.
}}
\label{figure:utaus}
\end{figure}

\begin{figure}
\vskip  2.0truecm
\includegraphics{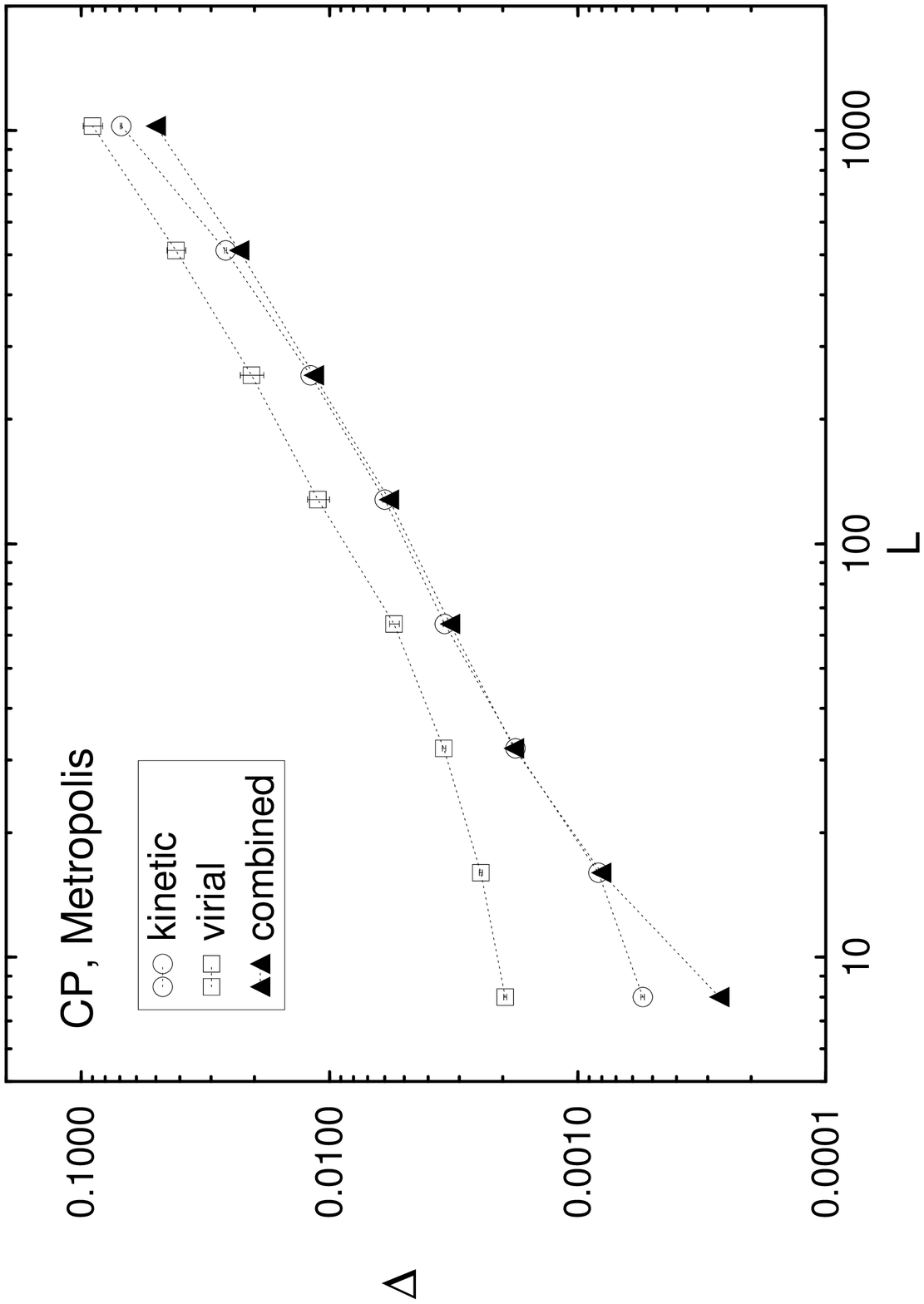}
\includegraphics{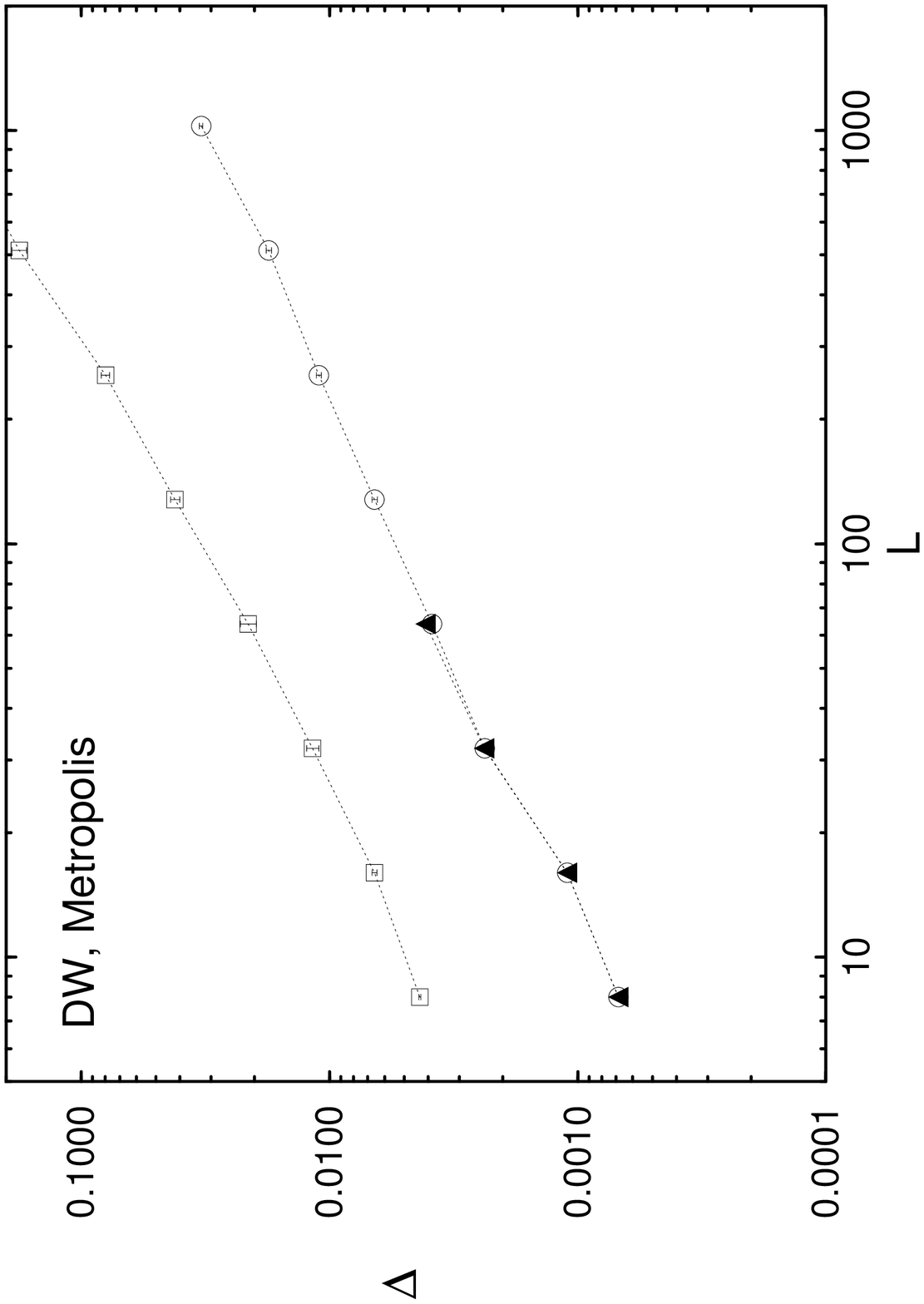}
\vskip 5.2truecm
\includegraphics{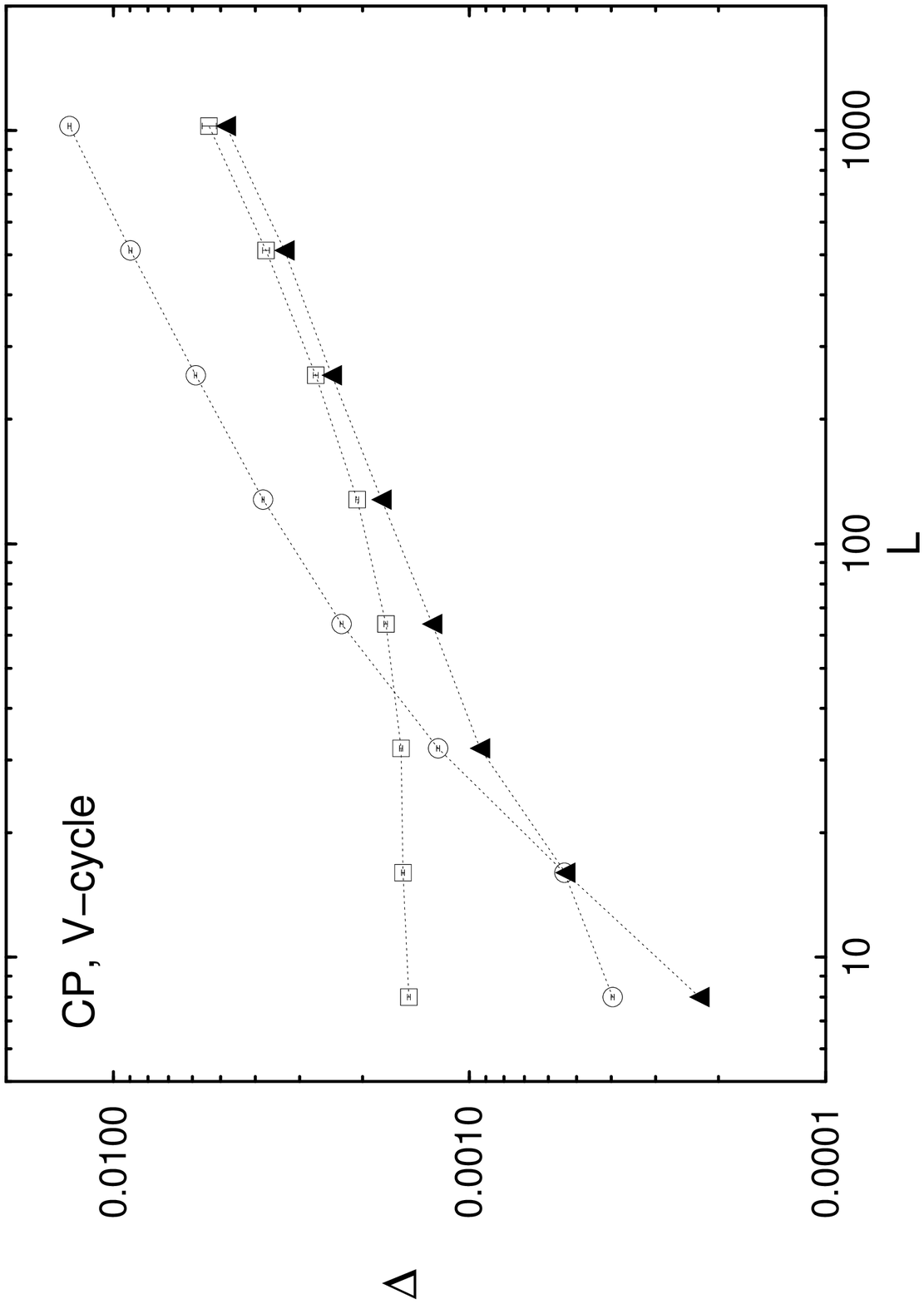}
\includegraphics{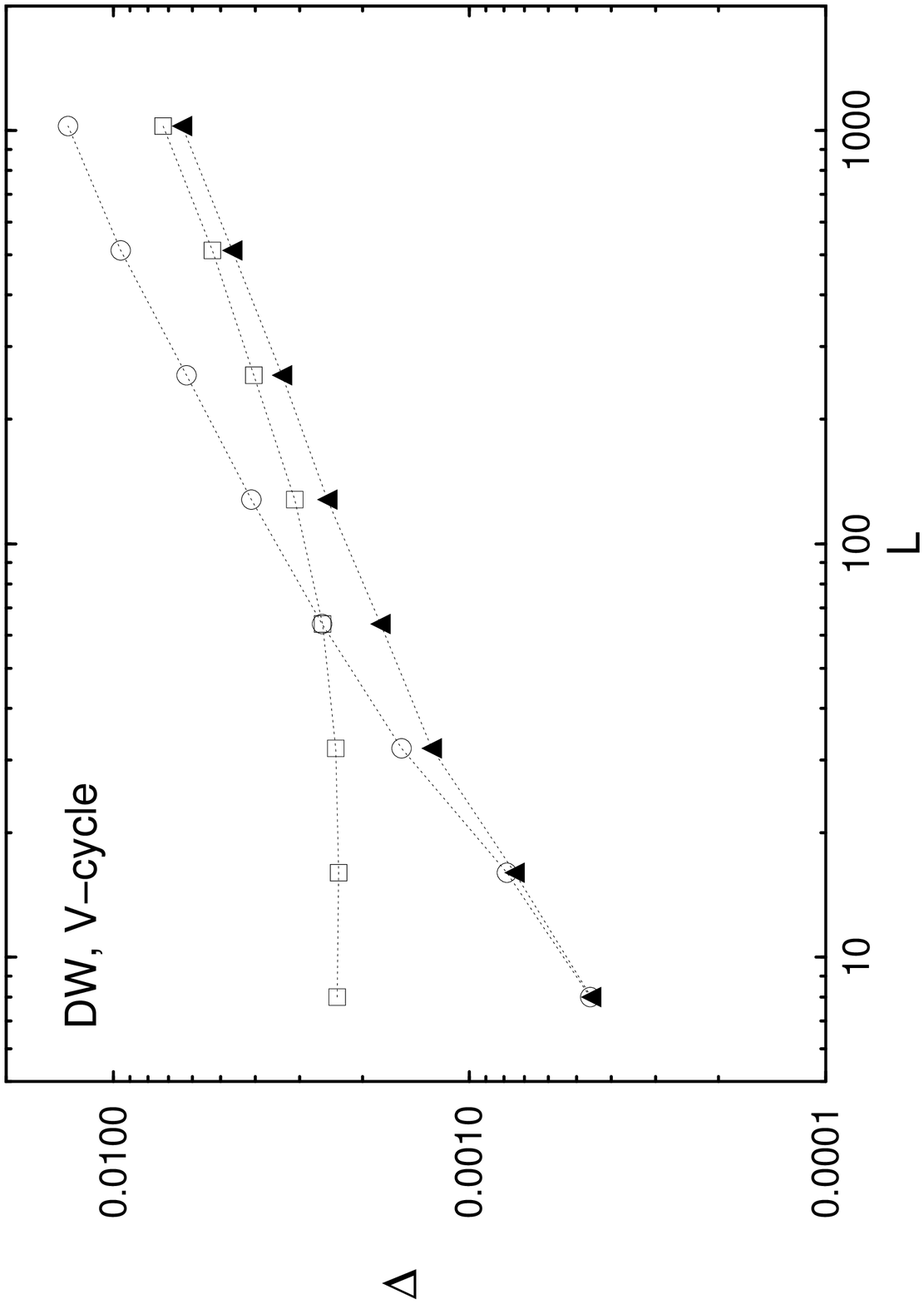}
\vskip 5.2truecm
\includegraphics{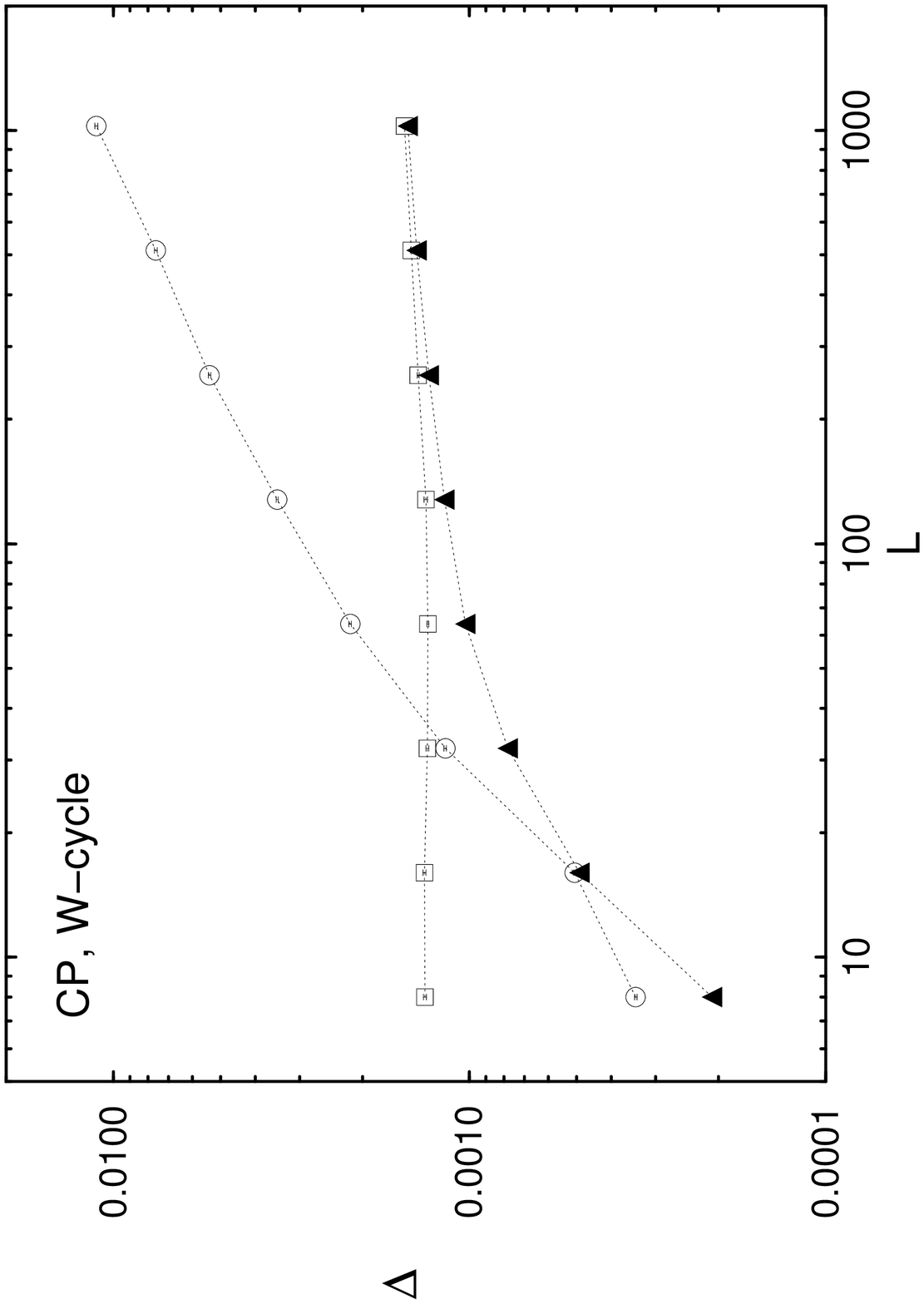}
\includegraphics{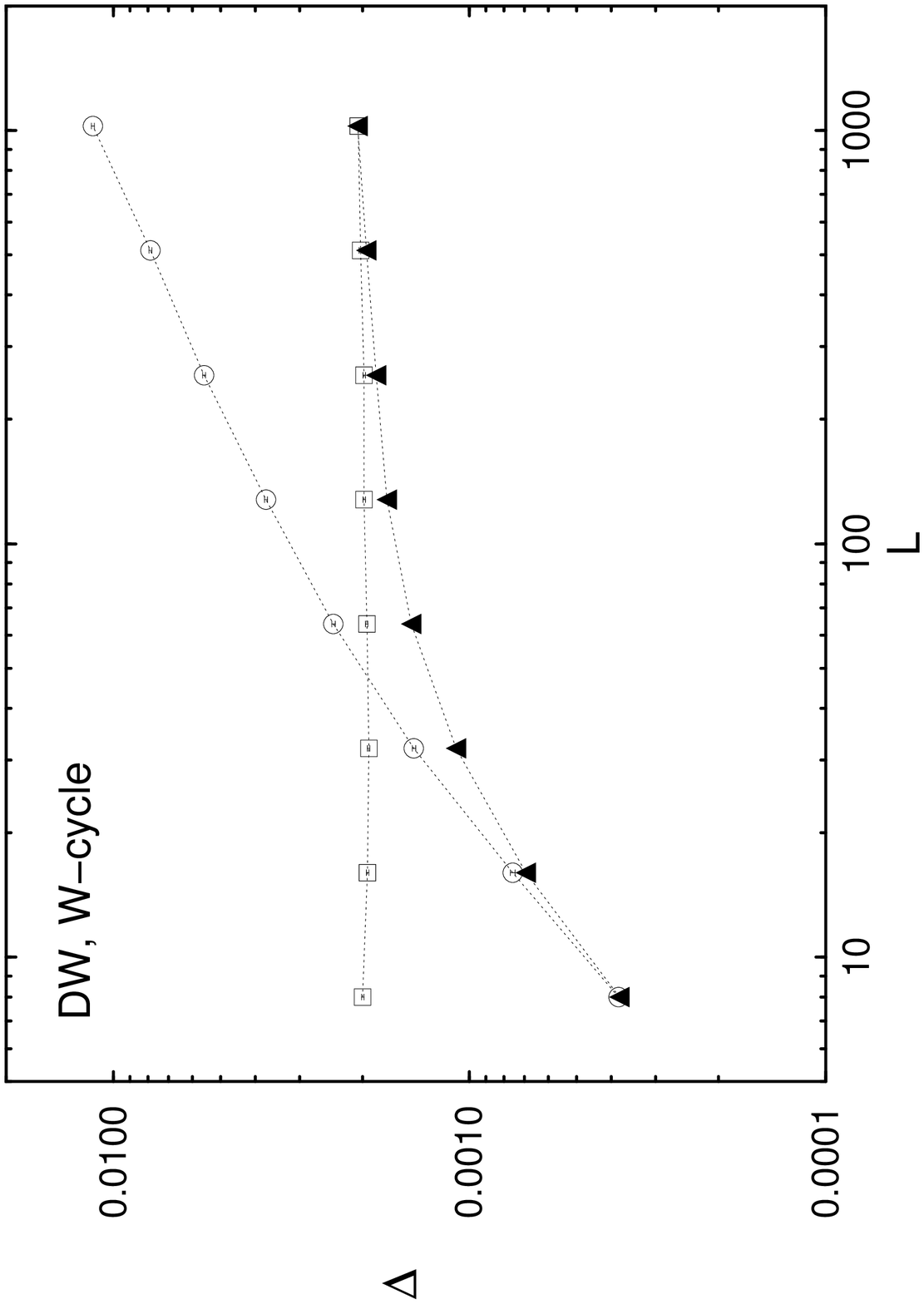}
\vskip 5.2truecm
\includegraphics{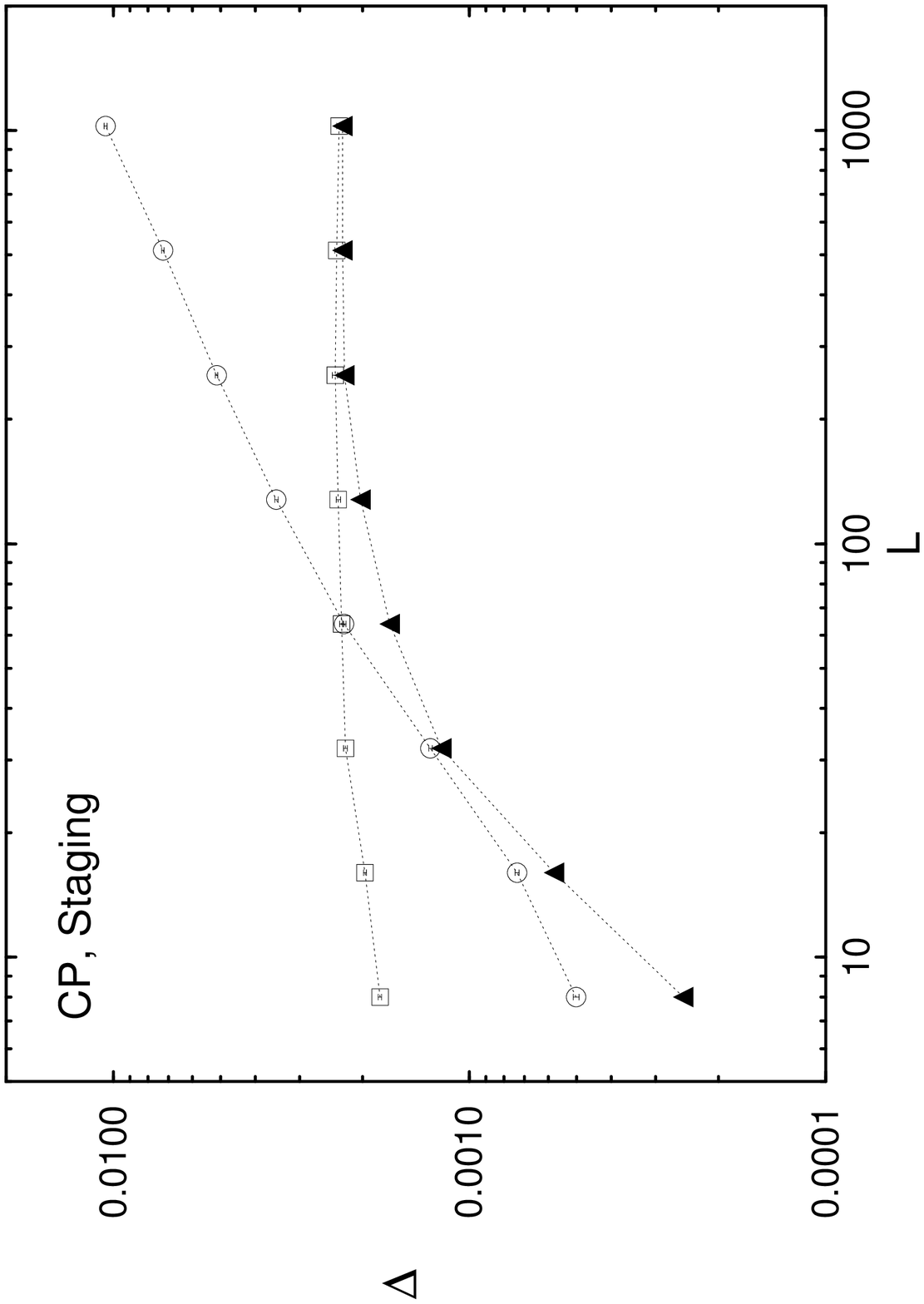}
\includegraphics{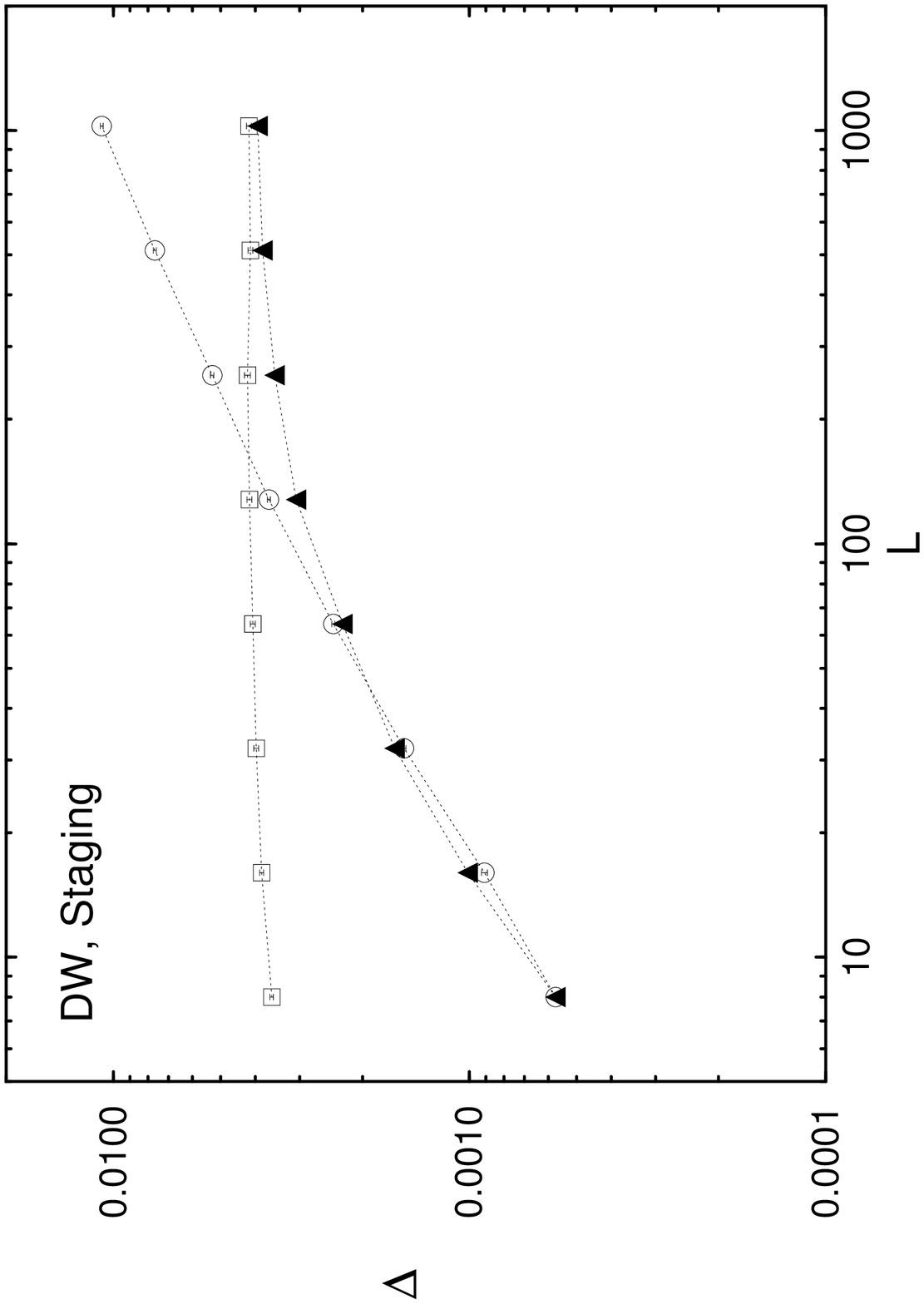}
\caption[a]{
The errors $\Delta \equiv \Delta \overline U$ 
computed from the variance and the autocorrelation
times for the three estimators using different update algorithms for
the convex potential (CP) and the double well (DW).
}
\label{figure:uerrs}
\end{figure}

The integrated autocorrelation times for the two potentials and the four
update algorithms collected in Tables
\ref{table:cptau} and \ref{table:dwtau} are plotted in 
Figs.~\ref{figure:utaus}(a)-(h) against $L$ on a double-logarithmic scale. 
As discussed in section
\ref{ssec:Discuss} and illustrated in Fig.~\ref{figure:sigma}, the
variance does not depend on the update algorithm but it strongly
depends on the estimator. As we will see, the different dependencies
of the variance and the autocorrelation times on the number of
variables $L$ combine in a peculiar way in 
the errors $\Delta \equiv \Delta \overline{U}$ which are plotted in
Figs.~\ref{figure:uerrs}(a)-(h).
Comparing the ``kinetic''
and the ``virial'' estimators, the combination often leads to a
crossover for the final error. The error of the combined estimator is,
however, always smaller than the better of the two estimators.
\paragraph{Metropolis algorithm:}
In Figs.~\ref{figure:utaus}(a) and (b) we observe for the 
autocorrelation time of the
virial
estimator the expected quadratic divergence. By fitting the power-law
ansatz $\tau_{\rm int}=\alpha L^z$ to the data for the convex potential (CP)
with $L=128$, $256$, $512$, and $1024$ we obtain indeed an exponent of
$z=2.017(13)$. 
The autocorrelation times
of the kinetic estimator in this case behave rather differently,
and at first glance it may seem that they
do not show the
expected $L^2$-divergence. It is clear, however, that the plotted fit with
an ``effective'' exponent of
$z=1.372(16)$ is not really justified since there is still distinct
curvature for the last three data points ($L=256$, $512$, $1024$). 
For
the double-well potential the difference between the 
two
estimators is even more pronounced. For the virial
estimator we obtain in the range $64 \le L \le 1024$ again an excellent fit 
with an exponent of $z=1.986(54)$, which is fully consistent with the
expected value of two. For the data of the kinetic estimator in the same 
range the quality of the fit is greatly reduced and the ``effective'' 
exponent of $z=0.461(15)$ is again found to be much smaller.  
By successively discarding small values of $L$ in the fit we observed 
a definite trend 
to larger values of $z$, but 
the available sizes of $L$ are still much too small to see the truly
asymptotic behaviour for the kinetic estimator.

We may, however, compare the autocorrelation times with those of the
moments $m_2\equiv\sum_{i=1}^L x_i^2$ and $m_4\equiv\sum_{i=1}^L
x_i^4$ reported in an earlier investigation~\cite{js93a},
which are of the same order of
magnitude. The same is true for the autocorrelation times of the
correlations $x_kx_{k-1}$ which we have elsewhere found~\cite{thesis}
for the convex potential to increase as $1.078(14)$, $1.903(22)$,
$4.466(94)$, $13.53(70)$, $52.3(6.6)$, and $227(21)$ for lattices of
size $L=8$, $16$, $32$, $64$, $128$, and $256$.  Surprisingly enough,
the autocorrelation times of the kinetic estimator are much smaller
and start to diverge only much later.  Recall that due to the periodic
boundary conditions and translational invariance the kinetic estimator
is a linear combination of $m_2$, $m_4$, and $x_kx_{k-1}$.  The
particular combination, however, has a much smaller autocorrelation
time.

As a consequence, contrary to previous expectations, for the
Metropolis algorithm the kinetic estimator turns out to be much more
accurate than the virial estimator for the parameters under
consideration. This is clearly seen in Figs.~\ref{figure:uerrs}(a) and
(b).
Note, however, that this result depends crucially on the
observed autocorrelation times in the range of $L$-values investigated
here.
If also for the kinetic
estimator the autocorrelation times asymptotically diverge as
$\tau_{\rm int,k} \propto L^2$, then we should find for very
large $L$ that 
$(\Delta \overline{U}_{\rm k})^2 \propto \sigma_{\rm k}^2 \tau_{\rm int,k} 
\propto L^3$, while the error for the virial estimator exhibits a
weaker $L$-dependence of $(\Delta \overline{U}_{\rm v})^2 \propto L^2$.

Since modified update algorithms such as the
staging algorithm and the multigrid W-cycle eliminate slowing down the
virial estimator for these update schemes will asymptotically {\em
always} be more favorable.

\paragraph{Multigrid V-cycle:}
The different behaviour of the autocorrelation times for the kinetic
and virial estimators is also found, albeit less pronounced, for the
V-cycle. Fitting our data for $L=256$, $512$, and $1024$ 
in Figs.~\ref{figure:utaus}(c) and (d)
we find for
the virial estimator exponents of $z=0.959(54)$ (CP) and $z=0.808(42)$
(DW), supporting the expected value of $z=1$. The corresponding
exponents for the moments $m_2$ of $z=0.8356(92)$ (CP) and
$z=0.715(27)$ (DW), obtained elsewhere \cite{js93a} fitting data for $L=128$,
$256$, and $512$, are compatible with these values taking into account
the fact that those data still showed some upward curvature and were
hence conjectured to underestimate the asymptotic behaviour.

Again, however, the kinetic estimator has much smaller autocorrelation
times. For $L \le 1024$ the fits give here effective exponents of
$z=0.135(18)$ (CP) and $z=0.078(23)$ (DW).  Even though we can of
course not exclude the possibility that the data again start to
diverge much faster for larger values of $L$, our data do not show any
tendency of upward curvature.

Taking into account the fact that also the variance for small values
of $L$ is much smaller for the kinetic estimator we find that, for
small $L$, the actual errors for the kinetic estimator are much
smaller as well. In Figs.~\ref{figure:uerrs}(c) and (d) we see that with 
increasing $L$ the difference decreases, and for
$L \approx 50$ both estimators would roughly yield the same error. For
large values of $L$ the virial estimator gets still better but for values
in the range $256 \leq L \leq 1024$ the errors for the two estimators
seem to increase with roughly the same effective exponent.

\paragraph{Multigrid W-cycle:}
The W-cycle exhibits a considerably improved dynamical behaviour and
the magnitude of the autocorrelation times 
shown in Figs.~\ref{figure:utaus}(e) and (f)
is now
greatly reduced to about unity. Notice that in contrast to all other
update algorithms here the virial estimator has smaller autocorrelations
than the kinetic estimator.
In particular the divergence with increasing $L$ is much weaker for both 
estimators. Only for the convex potential 
we still find a slight increase of the autocorrelation
times for the virial estimator with an exponent of $z=0.1087(65)$, while
the corresponding exponent
for the kinetic estimator, $z=0.052(11)$, is compatible with zero. 
For the double-well
potential both exponents ($z=0.040(63)$ for the virial and
$z=0.028(12)$ for the kinetic estimator) are consistent with zero.
Hence the continuum slowing down problem is solved for the
W-cycle. 
Again these estimates are in good agreement with the corresponding
exponents for the moments $m_2$ of $z=0.1043(29)$ (CP) and
$z=-0.015(11)$ (DW) obtained in our earlier investigation \cite{js93a}.

Since the autocorrelation times no longer diverge and since the
behaviour of the variances does not depend on the update algorithm, we
expect a crossover for the actual errors associated with the virial
and the kinetic estimator. Such a crossover was already observed for
the V-cycle in Figs.~\ref{figure:uerrs}(c) and (d),
but for the W-cycle in Figs.~\ref{figure:uerrs}(e) and (f) it is much more
pronounced. Basically this simply reflects the behaviour of the
variances $\sigma_{\rm v}^2$ and $\sigma_{\rm k}^2$ shown in
Fig.~\ref{figure:sigma}, which exhibit a clear crossover around $L=64$ for the
convex and around $L=100$ for the double-well potential, respectively. 
Since for large values of $L$ the statistical
errors of the virial
estimator remain roughly constant, it here always outperforms the
kinetic estimator whose errors increase due to the linear $L$-dependence 
of the variance.
\paragraph{Staging algorithm:}

From Figs.~\ref{figure:utaus}(g) and (h) 
it is obvious that also the staging algorithm eliminates slowing down.
Notice that for the staging algorithm the virial estimator has
larger autocorrelations than the kinetic estimator, while for the
W-cycle it is just the other way around.

It should be kept in mind, however, that the staging algorithm
requires the choice of an optimal staging length $j_{\rm opt}$. This problem 
was discussed elsewhere~\cite{js96a}
for the virial estimator
by explicitly looking at the autocorrelation
times as a function of the staging length.
It was shown that the optimal choice scales with the number of variables
$L$, and that the corresponding optimal acceptance rates stay roughly
constant for different $L$. Here we extend this discussion and
list in Table \ref{table:jopt} the values of $j_{\rm opt}$ 
and the corresponding acceptance rates
for both the virial and the kinetic estimator. We see that
the rule of thumb of some fixed acceptance
probability regardless of the estimators is rather
misleading.  
In fact, while for the virial estimator the optimal value of
$j$ corresponds to an acceptance rate of about 55\%, for the kinetic
estimator the optimum is at about $85\%-90\%$. 

It should be observed that for an optimal performance we had to use a
different $j_{\rm opt}$ for the two estimators, i.e., used data from
different simulations. For the convex potential and the largest grid
of $L=1024$ the kinetic estimator had the lowest autocorrelation time
of $\tau^{\rm opt}_{\rm int,k}=1.097(21)$ (cp.\ Table \ref{table:cptau}) 
for a
staging length of $j_{\rm opt} = 56$ with an acceptance rate of $90\%$.
Measuring the energy, on the other hand, for $j=176$ with an
acceptance rate of $56\%$, as was best for the virial estimator,
produced an autocorrelation time of $\tau_{\rm int,k}=1.483(37)$. The virial
estimator, on the other hand, had its lowest autocorrelation time of
$\tau^{\rm opt}_{\rm int,v}=2.48(56)$ for $j_{\rm opt}=176$. Here the
autocorrelation time is in fact more than doubled to a value of
$\tau_{\rm int,v}=4.53(17)$ if the optimal staging length $j=56$ for the 
kinetic estimator was used. This is nothing but yet another example of the
subtle interplay between update algorithm and energy estimator which
calls for some care when one aims at optimizing the efficiency of PIMC
simulations.

A bad choice of the staging length -- if only properly scaled with $L$
-- only affects the absolute value of the autocorrelation
times \cite{js96a}. If the staging length increases linearly with $L$
the corresponding autocorrelation times are roughly constant and do
not depend on $L$. This is confirmed by the data displayed in
Figs.~\ref{figure:utaus}(g) and (h) where we plotted fits with
exponents of $z=0.008(17)$ (CP) and $z=-0.005(20)$ (DW) for the virial
estimator, and with exponents of $z=-0.012(13)$ (CP) and $z=0.005(12)$
(DW) for the kinetic estimator.
All these exponents are fully consistent with zero and we conclude
that the staging algorithm eliminates slowing down just as well as the 
multigrid W-cycle. 
Looking at the error of the energy estimation in 
Figs.~\ref{figure:uerrs}(g) and (h),
which contains also the $L$-dependence of the variance, we hence
observe the very same crossover as we did for the W-cycle.

\subsection{The combined estimator}            \label{ssec:Res_combi}

If the choice between the two estimators were exclusive, one would
have to conclude that the best estimator would depend on details of
the simulation.
As discussed in section
\ref{ssec:Combi}, however, one can always combine the two estimators
and can hence always obtain errors which are even smaller than the
value for the better of the two estimators. Looking at the errors
displayed in Figs.~\ref{figure:uerrs}(a)-(h), one sees that for the convex
potential the gain
by combining the two estimators is best for very small values of $L$
where $\alpha$ is outside the range $[0,1]$ (cp. Tables
\ref{table:cpukvc} and \ref{table:dwukvc}). Another situation where
the combination appreciably reduces the error occurs, for both
potentials, if the actual
errors of the two estimators are of the same magnitude.

As was already suggested by Fig.~\ref{figure:interpol_cp}, for refined
update algorithms and large values of $L$ the optimal interpolation
parameter is always close to 0 which reflects the fact that the virial
estimator here always wins the race. For the Metropolis algorithm where no
crossover occurs, the opposite is true. Here the optimal $\alpha$ for
large $L$ is close to 1. For medium values of $L$, in the crossover
region for the V- and W-cycle and the staging algorithm, any values of
$\alpha$ were obtained.

Let us take a look at the autocorrelation times for the combined
estimator displayed in Figs.~\ref{figure:utaus}(a)-(h). For the
Metropolis algorithm the autocorrelation times (scaling effectively with 
exponents $z=1.413(21)$ (CP) and $z=0.469(51)$ (DW)) are close to but 
not necessarily smaller than those of
the kinetic estimator.
For the V-cycle the autocorrelation times of
the combined estimator ($z=0.523(38)$ (CP) and $z=0.388(33)$ (DW)) are
appreciably larger than those of the kinetic estimator but smaller
than those of the virial estimator. For the W-cycle ($z=0.136(12)$
(CP) and $z=0.0110 (88)$ (DW)) and for the staging algorithm
($z=0.021(17)$ (CP) and $z=0.042(17)$ (DW)) the situation is just
reversed. Here the combined autocorrelation times behave qualitatively
as the ones for the virial estimator.

Regardless whether the autocorrelation times of the combined estimator
are close to the smaller ones of either of the original estimators it
always yields errors which are slightly smaller. The gain is most
pronounced for very small lattices and in the crossover region.

The best gain in the reduction of the error by combining the
two estimators was obtained for $L=8$ and the Metropolis algorithm
where the error of the better estimator was in fact reduced by a
factor of 2. For $L=64$, the double well and the V-cycle the
errors of $0.31$ for the kinetic and $0.30$ for the virial estimator
were still reduced by a factor of 1.5 to an error of $0.21$.
Note that this reduction would be equivalent to a factor of more than
$2$ for the actual computer time needed to obtain the same reduction by
better statistics for the individual estimators.

Unfortunately, for most cases the gain by combining the estimators is
much smaller.  Realistically one may expect a gain of, say, $10\%$ in 
the final error by
the combination which may seem moderate a gain. It should be kept in
mind, however, that such a $10\%$ gain in the error corresponds to something
like a $20\%$ reduction of the computer run time needed to achieve the
same error by a simply increasing the statistics. And we emphasize
again that the reduction gained by combination of the estimators may
be obtained {\em after} the simulation completely without extra cost.

We finally observe that Fig.~\ref{figure:uerrs}(h) seems to contradict
our claim that the combination of the two estimators {\em always}
reduces the error of the better estimator. For the double well and the
staging algorithm with $L=16$ and $32$, the error of the
combined estimator indeed is larger than the error for the kinetic
estimator. The reason is that here the data for the kinetic estimator were
in fact obtained from the simulation optimized for the virial estimator. 
As pointed out above, the choice for the optimal staging length
depends on the estimator. The values for the two estimators $U_{\rm
k}$ and $U_{\rm v}$ reported in Tables \ref{table:cpukvc} and
\ref{table:dwukvc} were obtained for the respective {\em optimal}
staging length $j_{\rm opt}$ for each estimator (cp. Table
\ref{table:jopt}).  Clearly, the optimal staging length for the
combined estimator would be different both from the $j_{\rm opt}$ for
either the kinetic or the virial estimator. In fact, we are dealing
with an optimization problem of two dimensions in the parameter space
of $j_{\rm opt}$ and $\alpha$. However, in practical applications the
combination of the estimators would be done {\em after} the
simulation. Unfortunately, one has to decide beforehand which $j_{\rm
opt}$ best be chosen and regardless whether one might use $j_{\rm opt,
k}$ or $j_{\rm opt, v}$, in either case one would use a non-optimal
$j_{\rm opt}$ for one of the two estimators. In fact, if we
would have measured the energy using the kinetic estimator $U_{\rm k}$ 
using the optimal staging length for the virial
estimator, i.e. for $L=16$ with $j=6$ and for $L=32$ with $j=10$ we
would have obtained values of $-0.9564(11)$ and of $-0.9184(20)$
respectively. Clearly, these values are no longer superior to the ones
obtained using the combined estimator which were also computed for the
optimal $j$ for the virial estimator.

Thus, the apparent superiority of the kinetic estimator over the
combined estimator in Fig.\ref{figure:uerrs} (h) is in fact an
artifact of our over-careful data analysis.  In practice, when the
combination is done after the Monte Carlo run with data for the same
set of parameters (non-optimal for at least one of the estimators),
the combined estimator is guaranteed to yield the best energy
estimates with the smallest error bars.
%
                     \section{Conclusions}                \label{sec:conclusion}
%
Our concern in this paper was to show how energy estimation in PIMC
simulations can be optimized by taking into account both variances and
autocorrelation times of two standard energy estimators: the
``kinetic'' and the ``virial'' estimator, and by investigating their
respective interplay with different update algorithms: the standard
{\em local\/} Metropolis algorithm, the {\em non-local\/} multigrid V-
and W-cycles, and the {\em non-local} staging algorithm.

Let us briefly summarize the main points: 

\begin{itemize}

\item[(i)]
      While the variance of the virial estimator depends only very weakly
      on the discretization scale, the variance of the kinetic estimator
      diverges asymptotically according to $\sigma^2_{\rm k} = L/2\beta^2$.
      This behaviour is independent of the update algorithm.
 
\item[(ii)] The dynamics of the update algorithms affects the
      autocorrelation times of the standard estimators. For the
      Metropolis algorithm these diverge as $L^2$ but this behaviour
      can clearly be seen only for the virial estimator. In fact, the
      values of the autocorrelation times for the kinetic
      estimator turned out to be much smaller than those of the virial
      estimator, and we did not see the expected $L^2$ divergence with
      the parameters in our simulations. Refined non-local 
      update algorithms
      reduce (V-cycle) or eliminate (W-cycle, staging) an $L$
      divergence of the autocorrelation times. For the staging
      algorithm we observed quite a strong dependence of the
      autocorrelation times on the length of the staging segment.  In
      particular, we have shown that the optimal length is very
      sensitive to the choice of the energy estimator. For both
      estimators it scales, however, linearly with $L$. In terms of
      acceptance rates the best performance was obtained at about 55\%
      for the virial and about $85\%-90\%$ for the kinetic estimator,
      respectively.

\item[(iii)] The interplay between the variances of estimators and the
      dynamics of the update algorithms which affects the
      autocorrelation times turns out to be quite subtle. It
      furthermore also depends on the discretization parameter
      $L$. The kinetic estimator often has a smaller variance than the
      virial estimator for small $L$. Looking at the errors of the
      energy estimates we hence observe a crossover at which the
      virial estimator starts to win over the kinetic estimator since
      its variance does not increase with $L$. For the Metropolis
      algorithm, due to its small autocorrelation times for the
      kinetic estimator, the crossover point, however, is shifted to
      very large values of $L$ and was not seen with the parameters in
      our simulation.

\item[(iv)] As a simple solution to the involved interplay of the
      variance and the autocorrelation time for the two energy
      estimators we have introduced a ``combined estimator.'' By
      construction this {\em always} gives more accurate energy
      estimates than the better of the two standard estimators. The
      empirically observed gain varies but realistically one may
      expect a gain of about $10\%$ in the error for energy estimation
      in PIMC simulations. We emphasize, that this corresponds to a
      $20\%$ gain in actual simulation time which comes at no extra
      cost and is very easy to implement after the actual simulation.

\item[(v)] The combination of different estimators for the same
physical quantity is a very general option for Monte Carlo simulations
and by no means restricted to the use of the standard energy
estimators in PIMC simulations.

\end{itemize}

                      \section*{Acknowledgments}
W.J. thanks the Deutsche Forschungsgemeinschaft for a Heisenberg
fellowship.
%
                     
%
%
\clearpage
\scriptsize
\begin{table}[tb]
\catcode`?=\active \def?{\kern\digitwidth}
\centering
\begin{tabular}{|r|l|l|l|l|l|l|l|l|}
\hline
\multicolumn{9}{|c|}{Convex Potential (CP): $V(x) = \frac{1}{2}x^2+x^4$}\\
\hline & & & & & & & & \\
\multicolumn{1}{|c|}{$L$} &
\multicolumn{1}{c|}{$\Ubar_{\rm k}$}   &
\multicolumn{1}{c|}{\%}  &
\multicolumn{1}{c|}{$\Ubar_{\rm v}$}   &
\multicolumn{1}{c|}{\%}  &
\multicolumn{1}{c|}{$\rho$} &
\multicolumn{1}{c|}{$\alpha_{\rm opt}$} &
\multicolumn{1}{c|}{$\Ubar_{\rm c}$}   &
\multicolumn{1}{c|}{\%} \\
\hline
\multicolumn{9}{|c|}{Metropolis}\\
\hline
8    &  0.48883(57) & 0.12 & 0.4891(21) & 0.43 & 0.913(58) & 1.323 & 0.48874(29) & 0.060 \\
16   &  0.65741(86) & 0.13 & 0.6553(23) & 0.36 & 0.898(63) & 1.120 & 0.65766(83) & 0.13  \\
32   &  0.7534(18)  & 0.24 & 0.7552(31) & 0.42 & 0.86(11)  & 0.793 & 0.7537(17)  & 0.23  \\
64   &  0.7864(37)  & 0.47 & 0.7783(58) & 0.75 & 0.78(23)  & 0.825 & 0.7850(36)  & 0.46  \\
128  &  0.8021(67)  & 0.84 & 0.788(11)  & 1.4  & 0.67(42)  & 0.863 & 0.8001(66)  & 0.83  \\
256  &  0.8074(69)  & 0.86 & 0.791(11)  & 1.4  & 0.54(42)  & 0.933 & 0.8063(69)  & 0.86  \\
512  &  0.8169(73)  & 0.89 & 0.817(10)  & 1.3  & 0.42(39)  & 0.819 & 0.8169(72)  & 0.88  \\
1024 &  0.7990(96)  & 1.20 & 0.794(11)  & 1.4  & 0.30(38)  & 0.611 & 0.7973(86)  & 1.08  \\
\hline
\multicolumn{9}{|c|}{V-cycle}\\
\hline
8    &  0.48834(40) & 0.082 & 0.4871(16) & 0.33 & 0.912(49) & 1.253 & 0.48866(25) & 0.052 \\
16   &  0.65847(57) & 0.087 & 0.6581(16) & 0.24 & 0.898(62) & 0.969 & 0.65846(57) & 0.087 \\
32   &  0.7530(11)  & 0.15  & 0.7534(17) & 0.23 & 0.857(73) & 0.705 & 0.75313(81) & 0.11  \\
64   &  0.7916(23)  & 0.29  & 0.7864(18) & 0.23 & 0.78(14)  & 0.391 & 0.7885(14)  & 0.18  \\
128  &  0.8011(36)  & 0.45  & 0.7982(21) & 0.27 & 0.67(23)  & 0.261 & 0.7990(18)  & 0.23  \\
256  &  0.8085(58)  & 0.72  & 0.8051(26) & 0.33 & 0.54(35)  & 0.170 & 0.8057(24)  & 0.30  \\
512  &  0.8174(85)  & 1.04  & 0.8051(39) & 0.48 & 0.42(38)  & 0.192 & 0.8075(34)  & 0.42  \\
1024 &  0.807(13)   & 1.61  & 0.8008(56) & 0.70 & 0.31(45)  & 0.164 & 0.8018(52)  & 0.65  \\
\hline
\multicolumn{9}{|c|}{W-cycle}\\
\hline
8    &  0.48844(32) & 0.066 & 0.4882(13) & 0.27 & 0.912(42) & 1.228 & 0.48850(22) & 0.045 \\
16   &  0.65822(53) & 0.081 & 0.6589(13) & 0.20 & 0.898(59) & 0.855 & 0.65831(49) & 0.075 \\
32   &  0.7537(13)  & 0.18  & 0.7521(14) & 0.19 & 0.856(79) & 0.546 & 0.75297(88) & 0.12  \\
64   &  0.7885(21)  & 0.27  & 0.7904(15) & 0.19 & 0.78(12)  & 0.343 & 0.7898(12)  & 0.16  \\
128  &  0.7924(33)  & 0.42  & 0.7963(14) & 0.18 & 0.67(20)  & 0.192 & 0.7955(12)  & 0.15  \\
256  &  0.8053(53)  & 0.66  & 0.8045(14) & 0.17 & 0.54(30)  & 0.091 & 0.8045(13)  & 0.16  \\
512  &  0.8102(76)  & 0.94  & 0.8050(14) & 0.17 & 0.42(38)  & 0.032 & 0.8052(14)  & 0.17  \\
1024 &  0.801(11)   & 1.38  & 0.8033(18) & 0.22 & 0.31(39)  & 0.025 & 0.8032(18)  & 0.22  \\
\hline
\multicolumn{9}{|c|}{Staging}\\
\hline
8    &  0.48786(52) & 0.11 & 0.4861(19) & 0.39 & 0.913(53) & 1.337 & 0.48845(25) & 0.051 \\
16   &  0.65907(67) & 0.10 & 0.6594(20) & 0.30 & 0.899(69) & 1.121 & 0.65743(62) & 0.094 \\
32   &  0.7544(13)  & 0.17 & 0.7556(21) & 0.28 & 0.858(90) & 0.779 & 0.7546(12)  & 0.16  \\
64   &  0.7902(20)  & 0.25 & 0.7889(23) & 0.29 & 0.79(16)  & 0.411 & 0.7883(18)  & 0.23  \\
128  &  0.7958(38)  & 0.48 & 0.7980(22) & 0.28 & 0.67(26)  & 0.206 & 0.7990(19)  & 0.24  \\
256  &  0.8051(50)  & 0.62 & 0.8025(27) & 0.37 & 0.54(35)  & 0.101 & 0.8035(26)  & 0.32  \\
512  &  0.7981(71)  & 0.89 & 0.8050(26) & 0.32 & 0.41(48)  & 0.076 & 0.8035(25)  & 0.31  \\
1024 &  0.799(11)   & 1.4  & 0.7968(22) & 0.28 & 0.31(52)  & 0.047 & 0.7969(21)  & 0.26  \\
\hline
\end{tabular}
\caption[Measured energies using the kinetic and the virial estimators]
        {Convex Potential (CP): Measured energies using the kinetic
         estimator $U_{\rm k}$, the virial estimator $U_{\rm v}$,
         and the combined estimator $U_{\rm c}$.
         Also listed are the relative jacknife errors in percent, the
         cross-correlation coefficient $\rho$, and the parameter 
         $\alpha_{\rm opt}$ of the optimal combined estimator.
         }
\label{table:cpukvc}
\end{table}
\begin{table}[tb]
\catcode`?=\active \def?{\kern\digitwidth}
\centering
\begin{tabular}{|r|l|l|l|l|l|l|l|l|}
\hline
\multicolumn{9}{|c|}{Double Well (DW): $V(x) = -\frac{1}{2}x^2+0.04x^4$}\\
\hline & & & & & & & & \\
\multicolumn{1}{|c|}{$L$} &
\multicolumn{1}{c|}{$\Ubar_{\rm k}$}   &
\multicolumn{1}{c|}{\%} &
\multicolumn{1}{c|}{$\Ubar_{\rm v}$}   &
\multicolumn{1}{c|}{\%} &
\multicolumn{1}{c|}{$\rho$} &
\multicolumn{1}{c|}{$\alpha_{\rm opt}$} &
\multicolumn{1}{c|}{$\Ubar_{\rm c}$}   &
\multicolumn{1}{c|}{\%} \\
\hline
\multicolumn{9}{|c|}{Metropolis}\\
\hline
8    &  -1.04785(70) & 0.067 & -1.0516(36) & 0.34 & 0.88(13) & 1.000 & -1.04785(70) & 0.067 \\
16   &  -0.9547(12)  & 0.13  & -0.9576(72) & 0.75 & 0.84(32) & 0.991 & -0.9548(12)  & 0.13  \\
32   &  -0.9210(25)  & 0.28  & -0.916(12)  & 1.31 & 0.79(52) & 1.022 & -0.9211(25)  & 0.28  \\
64   &  -0.9051(31)  & 0.35  & -0.917(15)  & 1.64 & 0.73(56) & 0.943 & -0.9057(30)  & 0.91  \\
128  &  -0.8931(81)  & 0.91  & -0.905(12)  & 1.76 &          &       &              &       \\
256  &  -0.899(13)   & 1.42  & -0.916(11)  & 1.17 &          &       &              &       \\
512  &  -0.886(22)   & 2.42  & -0.916(15)  & 1.60 &          &       &              &       \\
1024 &  -0.889(19)   & 2.06  & -0.909(13)  & 1.38 &          &       &              &       \\
\hline
\multicolumn{9}{|c|}{V-cycle}\\
\hline
8    &  -1.04822(45) & 0.043 & -1.0484(26) & 0.25 & 0.88(10) & 1.012 & -1.04822(45) & 0.043 \\
16   &  -0.95403(81) & 0.085 & -0.9555(23) & 0.24 & 0.84(11) & 0.883 & -0.95421(76) & 0.080 \\
32   &  -0.9184(14)  & 0.15  & -0.9196(23) & 0.25 & 0.79(12) & 0.714 & -0.9187(12)  & 0.13  \\
64   &  -0.9073(28)  & 0.31  & -0.9080(27) & 0.30 & 0.73(16) & 0.488 & -0.9077(19)  & 0.21  \\
128  &  -0.9041(39)  & 0.44  & -0.9060(33) & 0.37 & 0.64(22) & 0.418 & -0.9052(25)  & 0.28  \\
256  &  -0.9108(65)  & 0.72  & -0.9053(38) & 0.42 & 0.53(31) & 0.248 & -0.9066(33)  & 0.37  \\
512  &  -0.9016(97)  & 1.08  & -0.9080(63) & 0.70 & 0.41(42) & 0.327 & -0.9059(48)  & 0.53  \\
1024 &  -0.902(13)   & 1.45  & -0.8918(80) & 0.90 & 0.31(47) & 0.274 & -0.8945(67)  & 0.75  \\
\hline
\multicolumn{9}{|c|}{W-cycle}\\
\hline
8    &  -1.04776(39) & 0.038 & -1.0479(20) & 0.19 & 0.876(81) & 1.020 & -1.04775(39) & 0.038 \\
16   &  -0.95539(79) & 0.083 & -0.9513(21) & 0.22 & 0.84(19)  & 0.872 & -0.95487(74) & 0.078 \\
32   &  -0.9179(15)  & 0.17  & -0.9195(19) & 0.21 & 0.79(12)  & 0.617 & -0.9185(12)  & 0.13  \\
64   &  -0.9104(29)  & 0.32  & -0.9107(21) & 0.23 & 0.73(13)  & 0.349 & -0.9106(16)  & 0.18  \\
128  &  -0.9054(34)  & 0.38  & -0.9045(21) & 0.24 & 0.64(18)  & 0.289 & -0.9047(17)  & 0.19  \\
256  &  -0.9036(57)  & 0.63  & -0.9039(21) & 0.24 & 0.53(23)  & 0.146 & -0.9038(18)  & 0.20  \\
512  &  -0.8979(75)  & 0.84  & -0.9001(22) & 0.25 & 0.41(30)  & 0.076 & -0.9000(22)  & 0.25  \\
1024 &  -0.8989(98)  & 1.09  & -0.9061(24) & 0.24 & 0.32(33)  & 0.073 & -0.9056(19)  & 0.21  \\
\hline
\multicolumn{9}{|c|}{Staging}\\
\hline
8    &  -1.04841(65) & 0.062 & -1.0542(37) & 0.35 & 0.88(13) & 1.003 & -1.04839(65) & 0.062 \\
16   &  -0.95318(91) & 0.095 & -0.9582(35) & 0.37 & 0.84(16) & 0.916 & -0.9566(11)  & 0.11 \\
32   &  -0.9189(15)  & 0.16  & -0.9157(44) & 0.48 & 0.79(20) & 0.817 & -0.9179(18)  & 0.20 \\
64   &  -0.9053(26)  & 0.29  & -0.9102(44) & 0.48 & 0.73(26) & 0.683 & -0.9081(25)  & 0.28 \\
128  &  -0.9013(37)  & 0.41  & -0.9086(45) & 0.50 & 0.64(27) & 0.490 & -0.9081(32)  & 0.35 \\
256  &  -0.9011(55)  & 0.61  & -0.8979(46) & 0.51 & 0.53(33) & 0.341 & -0.9005(37)  & 0.41 \\
512  &  -0.9178(80)  & 0.87  & -0.9031(42) & 0.47 & 0.42(35) & 0.229 & -0.9044(36)  & 0.40 \\
1024 &  -0.907(11)   & 1.2   & -0.8999(44) & 0.49 & 0.31(52) & 0.112 & -0.9004(41)  & 0.46 \\
\hline
\end{tabular}
\caption[Measured energies using the kinetic and the virial estimators]
        {Double Well (DW): Measured energies using the kinetic
         estimator $U_{\rm k}$, the virial estimator $U_{\rm v}$,
         and the combined estimator $U_{\rm c}$.
         Also listed are the relative jacknife errors in percent, the
         cross-correlation coefficient $\rho$, and the parameter 
         $\alpha_{\rm opt}$ of the optimal combined estimator.}
\label{table:dwukvc}
\end{table}
\begin{table}[tbc]
\catcode`?=\active \def?{\kern\digitwidth}
\centering
\begin{tabular}{|r|l|l|l|l|l|l|}
\hline
\multicolumn{7}{|c|}{Convex Potential (CP): $V(x) = \frac{1}{2}x^2+x^4$}\\
\hline
\multicolumn{1}{|c|}{$L$} &
\multicolumn{1}{c|}{$\sigma^2_{\rm k}$} &
\multicolumn{1}{c|}{$\tau_{\rm int,k}$}  &
\multicolumn{1}{c|}{$\sigma^2_{\rm v}$} &
\multicolumn{1}{c|}{$\tau_{\rm int,v}$}  &
\multicolumn{1}{c|}{$\sigma^2_{\rm c}$} &
\multicolumn{1}{c|}{$\tau_{\rm int,c}$}  \\
\hline
\multicolumn{7}{|c|}{Metropolis}\\
\hline
8    &  0.00916(13) & ?1.636(51) & 0.1049(14) & ???1.825(48) & 0.003440(28) & ?1.053(28) \\
16   &  0.01942(16) & ?1.758(51) & 0.1272(18) & ???2.378(56) & 0.02097(15)  & ?1.535(32) \\
32   &  0.07641(58) & ?2.079(45) & 0.1345(19) & ???4.44(14)  & 0.05683(53)  & ?2.879(79) \\
64   &  0.2249(17)  & ?2.619(81) & 0.1278(30) & ??11.8(1.0)  & 0.1569(13)   & ?3.40(14)  \\
128  &  0.5418(43)  & ?3.30(12)  & 0.1296(59) & ??47.6(9.6)  & 0.4055(34)   & ?4.07(17)  \\
256  &  1.1760(56)  & ?6.03(15)  & 0.1285(48) & ?165(36)     & 1.0247(49)   & ?6.47(17)  \\
512  &  2.469(12)   & 13.91(24)  & 0.1311(42) & ?660(110)    & 1.6593(83)   & 15.97(29)  \\
1024 &  5.015(23)   & 47.55(80)  & 0.1280(52) & 3160(570)    & 1.8888(83)   & 65.9(1.9)  \\
\hline
\multicolumn{7}{|c|}{V-cycle}\\
\hline
8    &  0.00929(10) & ?0.845(15) & 0.1059(11) & ???1.033(16) & 0.003658(32) & ?0.6988(72) \\
16   &  0.01965(24) & ?0.746(11) & 0.1297(20) & ???0.912(14) & 0.01978(26)  & ?0.732(11) \\
32   &  0.07632(46) & ?0.982(17) & 0.1335(10) & ???0.909(18) & 0.05284(35)  & ?0.824(13) \\
64   &  0.2248(11)  & ?1.163(22) & 0.1352(13) & ???1.092(22) & 0.08560(71)  & ?0.947(18) \\
128  &  0.5450(32)  & ?1.326(19) & 0.1350(12) & ???1.583(31) & 0.11181(83)  & ?1.399(25) \\
256  &  1.1847(72)  & ?1.460(25) & 0.1352(15) & ???2.701(82) & 0.1275(12)   & ?2.321(65) \\
512  &  2.451(13)   & ?1.638(30) & 0.1372(22) & ???5.07(21)  & 0.1785(16)   & ?3.062(90) \\
1024 &  5.014(28)   & ?1.758(32) & 0.1345(27) & ??10.80(94)  & 0.2299(24)   & ?5.07(25)  \\
\hline
\multicolumn{7}{|c|}{W-cycle}\\
\hline
8    &  0.009041(81) & ?0.6434(93) & 0.10449(96) & ???0.851(16) & 0.003796(26) & ?0.5702(36)\\
16   &  0.01962(12)  & ?0.652(11)  & 0.1292(12)  & ???0.692(12) & 0.02198(17)  & ?0.5485(69)\\
32   &  0.07590(44)  & ?0.902(12)  & 0.13406(93) & ???0.644(11) & 0.05400(36)  & ?0.5626(75)\\
64   &  0.2265(11)   & ?1.033(18)  & 0.1362(10)  & ???0.6278(66)& 0.08669(57)  & ?0.6098(72)\\
128  &  0.5454(29)   & ?1.103(16)  & 0.13335(91) & ???0.660(14) & 0.10754(73)  & ?0.641(12) \\
256  &  1.1794(73)   & ?1.222(19)  & 0.13640(85) & ???0.7119(93)& 0.12236(94)  & ?0.690(10) \\
512  &  2.460(13)    & ?1.177(22)  & 0.1357(11)  & ???0.781(12) & 0.12951(97)  & ?0.765(12) \\
1024 &  5.002(28)    & ?1.250(25)  & 0.1348(12)  & ???0.859(16) & 0.1314(12)   & ?0.845(14) \\
\hline
\multicolumn{7}{|c|}{Staging}\\
\hline
8    &  0.00900(11) & ?1.398(49) & 0.1032(12) & ???1.545(32) & 0.003454(26) & ?0.9084(36) \\  
16   &  0.01936(19) & ?1.396(35) & 0.1275(13) & ???1.518(24) & 0.02117(14)  & ?0.7878(39) \\
32   &  0.07628(45) & ?1.086(21) & 0.1347(16) & ???1.850(44) & 0.05588(42)  & ?1.287(32)  \\
64   &  0.2270(15)  & ?1.114(22) & 0.1336(14) & ???1.960(44) & 0.08546(75)  & ?1.644(41)  \\
128  &  0.5453(29)  & ?1.115(23) & 0.1357(16) & ???2.012(49) & 0.1083(12)   & ?1.885(47)  \\
256  &  1.1881(66)  & ?1.110(18) & 0.1344(18) & ???2.119(67) & 0.1209(15)   & ?2.086(64)  \\
512  &  2.448(13)   & ?1.076(16) & 0.1339(16) & ???2.079(53) & 0.1281(15)   & ?2.019(57)  \\
1024 &  5.054(29)   & ?1.097(21) & 0.1319(17) & ???2.048(56) & 0.1306(16)   & ?1.983(54)  \\
\hline
\end{tabular}
\caption[Measured energies using the kinetic and the virial estimators]
        {Convex Potential (CP): Variances $\sigma^2_{\rm k,v,c}$
         and the integrated autocorrelation times
         $\tau_{\rm int,k,v,c}$ for the kinetic (k),
         virial (v) and combined (c) energy estimator.
         }
\label{table:cptau}
\end{table}
\begin{table}[tb]
\catcode`?=\active \def?{\kern\digitwidth}
\centering
\begin{tabular}{|r|l|l|l|l|l|l|}
\hline
\multicolumn{7}{|c|}{Double Well (DW): $V(x) = -\frac{1}{2}x^2+0.04x^4$}\\
\hline
\multicolumn{1}{|c|}{$L$} &
\multicolumn{1}{c|}{$\sigma^2_{\rm k}$} &
\multicolumn{1}{c|}{$\tau_{\rm int,k}$}  &
\multicolumn{1}{c|}{$\sigma^2_{\rm v}$} &
\multicolumn{1}{c|}{$\tau_{\rm int,v}$}  &
\multicolumn{1}{c|}{$\sigma^2_{\rm c}$} &
\multicolumn{1}{c|}{$\tau_{\rm int,c}$}  \\
\hline
\multicolumn{7}{|c|}{Metropolis}\\
\hline
8    &  0.01282(30) & ?1.833(76) & 0.3645(40) & ????2.564(48) & 0.00128(30) & 1.834(76) \\
16   &  0.03257(41) & ?1.87(13)  & 0.3683(74) & ????5.88(24)  & 0.00322(41) & 1.89(13)  \\
32   &  0.1050(14)  & ?2.68(20)  & 0.364(12)  & ???18.7(2.1)  & 0.1096(14)  & 2.59(19)  \\
64   &  0.2598(19)  & ?3.206(82) & 0.352(13)  & ???64.4(9.4)  & 0.2327(17)  & 3.62(11)  \\
128  &  0.5746(42)  & ?3.75(20)  & 0.361(11)  & ??243(21)     &             &           \\
256  &  1.217(11)   & ?4.99(24)  & 0.3500(95) & ??909(67)     &             &           \\ 
512  &  2.498(19)   & ?6.17(33)  & 0.358(13)  & ?4407(570)    &             &           \\
1024 &  5.048(23)   & 10.68(34)  & 0.365(12)  & 14752(1600)   &             &           \\
\hline
\multicolumn{7}{|c|}{V-cycle}\\
\hline
8    &  0.01247(18) & ?0.839(53) & 0.3682(28) & ????0.752(18) & 0.00124(19) & 0.840(50) \\
16   &  0.03356(27) & ?0.918(17) & 0.3636(29) & ????0.746(12) & 0.00330(23) & 0.842(13) \\
32   &  0.10508(68) & ?1.146(23) & 0.3590(27) & ????0.784(12) & 0.00840(57) & 0.968(15) \\
64   &  0.2605(16)  & ?1.290(27) & 0.3538(26) & ????0.945(16) & 0.1551(11)  & 1.021(18) \\
128  &  0.5830(35)  & ?1.438(28) & 0.3575(36) & ????1.338(25) & 0.2250(17)  & 1.396(27) \\
256  &  1.2168(65)  & ?1.598(37) & 0.3550(41) & ????2.297(61) & 0.2749(28)  & 2.057(57) \\
512  &  2.519(14)   & ?1.814(41) & 0.3561(58) & ????3.92(15)  & 0.4293(39)  & 2.498(65) \\
1024 &  5.037(30)   & ?1.790(38) & 0.3657(74) & ????7.19(43)  & 0.5731(65)  & 3.60(14)  \\
\hline
\multicolumn{7}{|c|}{W-cycle}\\
\hline
8    &  0.01215(13) & ?0.597(20) & 0.3631(24) & ????0.5478(76) & 0.00120(13) & 0.598(20)  \\
16   &  0.03415(24) & ?0.836(26) & 0.3647(25) & ????0.5121(79) & 0.00340(22) & 0.709(22)  \\
32   &  0.10389(63) & ?0.987(20) & 0.3587(23) & ????0.5128(74) & 0.00930(61) & 0.639(13)  \\
64   &  0.2618(14)  & ?1.110(27) & 0.3573(25) & ????0.5291(75) & 0.1837(13)  & 0.5762(91) \\
128  &  0.5856(33)  & ?1.191(22) & 0.3630(28) & ????0.5377(72) & 0.2323(17)  & 0.6266(88) \\
256  &  1.2157(67)  & ?1.274(22) & 0.3566(26) & ????0.5479(72) & 0.2864(20)  & 0.5841(74) \\
512  &  2.504(14)   & ?1.236(24) & 0.3627(26) & ????0.5650(95) & 0.3246(24)  & 0.5842(92) \\
1024 &  5.070(28)   & ?1.287(23) & 0.3564(24) & ????0.5920(76) & 0.3345(22)  & 0.6360(84) \\
\hline
\multicolumn{7}{|c|}{Staging}\\
\hline
8    &  0.01242(30) & ?1.323(29) & 0.3631(38) & ????1.779(36) & 0.01239(31) & 1.325(29) \\
16   &  0.03330(34) & ?1.240(48) & 0.3582(41) & ????2.059(44) & 0.03242(43) & 1.579(61) \\
32   &  0.10403(76) & ?1.120(23) & 0.3628(49) & ????2.182(63) & 0.08351(77) & 1.575(43) \\
64   &  0.2606(16)  & ?1.117(22) & 0.3539(48) & ????2.335(64) & 0.1568(14)  & 1.648(27) \\
128  &  0.5847(35)  & ?1.145(23) & 0.3582(49) & ????2.406(76) & 0.2337(25)  & 2.017(49) \\
256  &  1.2242(67)  & ?1.139(24) & 0.3597(52) & ????2.456(87) & 0.2980(32)  & 2.085(71) \\
512  &  2.520(14)   & ?1.164(20) & 0.3592(52) & ????2.366(64) & 0.3473(36)  & 2.095(49) \\
1024 &  5.067(27)   & ?1.150(20) & 0.3610(49) & ????2.406(69) & 0.3478(42)  & 2.224(59) \\
\hline
\end{tabular}
\caption[Measured energies using the kinetic and the virial estimators]
        {Double Well (DW): Variances $\sigma^2_{\rm k,v,c}$
         and the integrated autocorrelation times
         $\tau_{\rm int,k,v,c}$ for the kinetic (k),
         virial (v) and combined (c) energy estimator.
         }
\label{table:dwtau}
\end{table}
\normalsize

\begin{table}[tb]
\catcode`?=\active \def?{\kern\digitwidth}
\centering
\vspace*{0.3cm}
\begin{tabular}{|r|r|l|r|l|}
\hline
\multicolumn{1}{|c}{$L$}              &
\multicolumn{1}{|c}{$j_{\rm opt,k}$}  & 
\multicolumn{1}{|c}{\%}               &
\multicolumn{1}{|c}{$j_{\rm opt,v}$}  & 
\multicolumn{1}{|c|}{\%}               \\
\hline
\multicolumn{5}{|c|}{Convex Potential (CP)}\\
\hline
8    &  2 & 64 &   2 & 64  \\
16   &  4 & 49 &   2 & 82  \\
32   &  4 & 73 &   4 & 72  \\
64   &  4 & 90 &  10 & 68  \\
128  & 10 & 86 &  24 & 55  \\
256  & 16 & 90 &  44 & 55  \\
512  & 24 & 93 &  88 & 55  \\
1024 & 56 & 90 & 176 & 56  \\
\hline
\multicolumn{5}{|c|}{Double-Well (DW)}\\
\hline
8    &   2 & 71 &   2 & 71 \\
16   &   4 & 63 &   6 & 47 \\
32   &   4 & 80 &  10 & 49 \\
64   &   8 & 84 &  20 & 52 \\
128  &  16 & 80 &  40 & 54 \\ 
256  &  32 & 82 &  80 & 54 \\
512  &  64 & 84 & 160 & 54 \\
1024 & 128 & 84 & 320 & 54 \\
\hline
\end{tabular}
\caption[a]{Optimal staging lengths $j_{\rm opt,k}$, $j_{\rm opt,v}$   and
         acceptance rates for the kinetic (k) and the virial (v) estimators.}
\label{table:jopt}
\end{table}
\clearpage
%
%
\newpage
%
\clearpage
{\Large\bf Figure Captions}
%
  \vspace{1cm}
  \begin{description}
    \item[\tt\bf Fig. 1:] 
  Variance of the individual energy measurements using the ``kinetic''
  and the ``virial'' estimators for the two potentials (\ref{eq:CP}) and
  (\ref{eq:DW}) at $\beta=10$. While the variance
  of the virial estimator is roughly constant in the continuum limit
  $L \longrightarrow \infty$, the variance of the
  kinetic estimator asymptotically diverges as $\sigma_{\rm k}^2 = 
  L/2\beta^2$.
    \item[\tt\bf Fig. 2:] 
  The optimal interpolation parameter $\alpha_{\rm opt}$ as given in
  eq.~(\ref{eq:alpha_kappa}) as a function of the ratio 
$\kappa = \Delta_{\rm v}/\Delta_{\rm k}$ and the correlation coefficient 
$\rho=\Delta_{\rm kv}^2/(\Delta_{\rm k} \Delta_{\rm v})$. 
    \item[\tt\bf Fig. 3:] 
    The reduction factor $R$ as given in eq.~(\ref{eq:R}) as a function
    of the ratio 
$\kappa = \Delta_{\rm v}/\Delta_{\rm k}$ and the correlation coefficient 
$\rho=\Delta_{\rm kv}^2/(\Delta_{\rm k} \Delta_{\rm v})$.
    \item[\tt\bf Fig. 4:] 
    Relative statistical error 
    $\Delta \overline{U}_{\rm c}/\overline{U}_{\rm c}$ 
    of the linear combination 
    $\overline{U}_{\rm c}$ 
    of the two energy estimators for the convex potential as a 
    function of the interpolation parameter $\alpha$ (using the W-cycle
    update algorithm). Data symbols
    denote jackknife averages over $100$ blocks, and the solid lines 
    are computed according to the theoretical prediction 
    (\ref{eq:sigmacombinedgeneral}), using as input only the (jackknife) 
    variances 
    of the virial and the kinetic estimator (corresponding to 
    $\alpha=0$, resp. $\alpha=1$) and the (jackknife) covariance of
    the two estimators. The arrows indicate the values of optimal 
    $\alpha$ according to eq.~(\ref{eq:alphaopti}).
    \item[\tt\bf Fig. 5:] 
(a) The autocorrelation function $A(j)$ on a logarithmic scale and (b) the
integrated autocorrelation time $\tint(k)$ of the virial estimator as
obtained with the V-cycle multigrid algorithm for the convex potential
(CP) at $\beta=10$ and $L=512$. Solid lines show fits according to
eq.~(\ref{eq:Afit}) resp. (\ref{eq:Bfit}). The asymptotic value of
$\tint(k)$ quoted in Table~\ref{table:cptau} is $\tau_{\rm int,v} = 5.07(21)$.
    \item[\tt\bf Fig. 6:] 
Integrated autocorrelation times $\tau_{\rm int}$ on a logarithmic scale 
for the three energy estimators
using different update algorithms for the convex potential (CP)
and the double well (DW).
Straight lines show fits of the form $\tau_{\rm int} = \alpha L^z$.
    \item[\tt\bf Fig. 7:] 
The errors $\Delta \equiv \Delta \overline U$ 
computed from the variance and the autocorrelation
times for the three estimators using different update algorithms for
the convex potential (CP) and the double well (DW).
\end{description}

\begin{thebibliography}{19}
\bibitem{fn1} All operators will be denoted by a ``hat'', so that it
  is easy to identify whether a quantum statistical average $\langle
  \dots \rangle$ is to be understood in the operator or path-integral
  sense.
\bibitem{fn2}
Note that even though $\la \frac{1}{L} \sum_{k=1}^L f(x_k) \ra
= \la f(x_k) \ra$ is trivially true for any function $f(x)$ due to the
periodic boundary conditions in the path integral (\ref{eq:Z_PI_dis}),
the proper definition of estimators should always contain the summation
over the path. This becomes crucial when considering the variance of the
estimator which involves a term
$\la \left(\frac{1}{L} \sum_{k=1}^L f(x_k) \right)^2 \ra \ne
\la f(x_k)^2 \ra$.
\bibitem{fn3} We would still have a valid energy estimator, though, as
  for any choice of $\alpha$.
\bibitem{fn5} Recall that for the Metropolis algorithm measurements
  were performed only every $n_e$ sweep. Here we have rescaled the
  measured autocorrelation times to units of sweeps.
\bibitem{fn6} For a comparison of Figs.~\ref{figure:uerrs}(a) and (b)
  with Tables \ref{table:cpukvc} and \ref{table:dwukvc} one has to
  keep in mind that the errors given in the tables refer to the full
  simulation with $n_e \times N_m$ sweeps, while in the figures we
  have rescaled them to hypothetical runs with $N_m=100\,000$ sweeps
  (=measurements), i.e., the numbers in the tables have to multiplied
  with $\sqrt{n_e}$ to obtain the data points in the figures.
\bibitem{fn7} This is, in fact, what we expect on general theoretical
  grounds.
\bibitem{fn8} For the Metropolis algorithm the autocorrelation times
  of the two estimators differ so much for $L \ge 128$ that it makes
  no sense to compute the combined estimator.
\end{thebibliography}
\end{document}